\documentclass[12pt]{article}
\usepackage[utf8]{inputenc}
\usepackage[ inner=0.8cm, outer=0.8cm, top=1cm, bottom=1.9cm]{geometry}
\usepackage[svgnames]{xcolor}
\usepackage{listings}

\usepackage[bottom]{footmisc}
\usepackage{braket}
\usepackage{graphicx}
\usepackage{framed}
\usepackage{csquotes}
\usepackage{tikz}
\usetikzlibrary{decorations.markings}
\usetikzlibrary{decorations.pathmorphing}
\definecolor{shadecolor}{rgb}{0.90,0.90,0.90}
\usepackage{hyperref}
\usepackage{subcaption}
\usepackage{pifont}
\usepackage{setspace}
\usepackage{amsmath, amssymb, amsthm, float, graphicx,amsfonts}
\numberwithin{equation}{section}

\hypersetup{colorlinks=true, linkcolor=DarkRed, citecolor=DarkRed,urlcolor=DarkGreen,linktocpage}
\def\beq{\begin{eqnarray}}\def\eeq{\end{eqnarray}}
\def\be{\begin{equation}}\def\ee{\end{equation}}
\def\a{\alpha}
\def\b{\beta}

\def\d{\delta}

\def\g{\gamma}

\def\i{\iota}

\def\l{\lambda}

\def\r{\rho}
\def\s{\sigma}

\def\w{\omega}

\def\pd{\partial}

\def\la{\langle}
\def\ra{\rangle}

\def\tr{{\rm tr~}}

\usepackage{bm}
\usepackage{bm}

\begin{document}
\thispagestyle{empty}
\hfill
\vspace{2cm}
\begin{center}
{\LARGE\bf Exploring $2d$ localization with a step dependent coin}\\
\bigskip\vspace{0.5cm}{
{\large Kallol Sen}\let\thefootnote\relax\footnotetext{email: kallolmax@gmail.com} 
} \\[5mm]
{\it ICTP South American Institute for Fundamental Research,\\
\it  IFT-UNESP (${\it 1}^\circ$ andar), Rua Dr. Bento Teobaldo Ferraz 271, Bloco 2 - Barra Funda, \\
\it 01140-070 S\~{a}o Paulo, SP Brazil}\\
 \end{center}
\bigskip
\centerline{\large\bf Abstract}

\begin{quote}\small 
We generalize the coin operator of \cite{Zahed_2023}, to include a step dependent feature which induces localization in $2d$. This is evident from the probability distributions which can  be further used to categorize the localized walks. Localization is also evident from the entropic measures. We compute and compare three distinct measures (a) Shannon Entropy in the position and coin space, (b) Entanglement entropy between position and spin space, and (c) Quantum Relative Entropy which is a POVM of density operators of the step dependent and step independent coins. Shannon Entropy and Entanglement Entropy are periodic and bounded functions of the time steps. The zeros of Shannon and Entanglement entropies signify a complete localization of the wave-function. The Quantum Relative Entropy and Quantum Information Variance exhibit a similar periodic feature with a zero minima where the step dependent and step independent walks coincide. Finally, we compute the numerical localization length (inverse of the Lyapunov Exponent) for the step dependent coin as a function of energy and compare with an approximate perturbative computation, where we put the step dependent coin as a perturbation in the background of a step independent coin. In both the instances, we find that the localization length peaks at approximately the same positions in the momentum space.    
\end{quote}

\tableofcontents

\onehalfspacing
\newpage

\section{Introduction}
Localization in physics \cite{Romanelli_2005, MacKinnon_1983}, describes a phenomenon where a disorder or decoherence in the system can give rise to a localized wave function within the system. Localization has been in study for its interesting and important implications regarding  quantum transport \cite{Karamlou:2021dqa}, critical exponents \cite{Slevin_2014} and even in supersymmetric theories \cite{Pestun_2017}. The transition from weak to strong localization, also called Anderson transition, a field of active study in disordered spin chains \cite{PhysRevLett.127.230603, PhysRevA.99.060101, PhysRevE.104.054105}. 
In this work, we study the decoherence effects of a step dependent coin on random walks following the $1d$ approach of \cite{Panahiyan_2018}. Several aspects of localization in random walk has been considered in \cite{cmp/1103940927} for random environments, for entropy guided walks \cite{PhysRevLett.102.160602} and in the effects of recurrence \cite{Segawa:2013:1546-1955:1583} and so on. In this work we consider the simple generalization of the approach in \cite{Zahed_2023} and extend the analysis of \cite{Panahiyan_2018, Vakulchyk_2017}, to show that a step dependent coin is capable of inducing entanglement in the random walk in $2d$. The step dependent coin is given by,
\begin{shaded}
\be
\mathbf{C}_t(\theta_1,\theta_2,\phi)=\left(\begin{array}{cccc}e^{-it\phi}\cos t\theta_1&0&0&-ie^{-it\phi}\sin t\theta_1\\0& e^{it\phi}\cos t\theta_2&-ie^{it\phi}\sin t\theta_2&0\\0& -ie^{it\phi}\sin t\theta_2&e^{it\phi}\cos t\theta_2&0\\-ie^{-it\phi}\sin t\theta_1&0&0&e^{-it\phi}\cos t\theta_1\end{array}\right)\,.
\ee
\end{shaded}
that is capable of inducing localization in the walk. The corresponding step independent coin is given by $\mathbf{C}_{SIC}=\mathbf{C}_{t=1}$. At each step, we also compare the walks between the  step dependent coin (SDC) and its corresponding step independent coin (SIC) counterpart. We summarize our main findings below.
\begin{shaded}
\begin{itemize}
\item Since a non-zero probability distribution for our specific form of coin exists only along the diagonals, we compare SDC and SIC walks with $\theta_1=\theta_2=\theta$ for a symmetric localization. For illustration, we consider $\theta=\pi/(4j), \pi/(4j-1)$ for $j=1,2,3,\dots$ and $\phi=0$ and an initial wave-function at the origin. For $\theta=\pi/(4j)$, as $j$ increases, the SDC distribution localizes at the origin while the SIC walk spreads out. For $\theta=\pi/(4j-1)$, the SDC and SIC walks are diffused (and along the diagonals) for $j=1$. However, as $j$ increases, the distribution for the SDC walk becomes more pronounced around the origin. Consequently, for $\theta=\pi/(4j)$, the number of points with non-zero probability for SDC walk follows a periodic profile with an upper bound, while the SIC walk grows linearly with time steps. For $\theta=\pi/(4j-1)$, the number of points with non-zero probability starts off as an increasing function of time steps for $j=1$. However as $j$ increases, the number of points with non-zero probability decrease in agreement with the probability distribution. We also demonstrate that starting with a quantum walk for $\theta=\pi/3(1+j/10)$, we can transition to various regimes ranging from compact quantum walk to a semi-classical compact walk to a fully localized domain with various $1\leq j\leq 10$. 
\item Localization is also evident from dynamical observables point of view. The relation between localization and entanglement has been studied extensively either in random walk perspective of entanglement induced localization \cite{Crespi_2013} or their interconnections in non-relativistic QFTs \cite{Yngvason_2015, Rau_2003}. For our purposes, we explore the implications of entropic measures as an evidence for localization. We will use three different entropic variables (a) Shannon Entropy  \cite{Panahiyan_2018}, (b) Entanglement Entropy \cite{Zahed_2023} and (c) Quantum Relative Entropy \cite{Leditzky,Leditzky_2016,Leditzky_20161}. The measures (a) and (b) are state dependent while (c) is a POVM measurement. We observe that in comparison to the SIC where the Shannon entropy grows with time steps and the entanglement is an oscillatory function around a non-zero mean value, the SDC Shannon entropy and entanglement are oscillates with a lower bound at zero. In particular the zeros of the Shannon and entanglement entropy denote a separable state hinting at a completely localized state. For (c) we directly compare the density operators for the SDC and SIC walks using Quantum Relative Entropy (QRE) and Quantum Information Variance (QIV). These measures follow a periodic pattern between maxima and minima. The maxima signifies that the walks are most distinct from each other and vice-versa interpretation for the minima. At certain time steps, the variance is zero and coincides with the zeroes of QRE as well. The maxima denotes maximal localization for SDC, while the minima signifies maximum diffusion. 
\item We characterize the SDC walk by its {\it Localization Length} ($L_{loc}$), which implies the extent of the region within which the walk is bounded. The localization length is inverse of the {\it Lyapunov Exponent} ($\l_t$) which characterizes the strength of disorder (chaos) in the system. We follow the transfer matrix approach of \cite{Vakulchyk_2017}. For large disorder, the $L_{loc}$ is sufficiently small indicating strong localization. For $\theta=\pi/(4j)$, the walk is completely localized indicating $L_{loc}=0$. This is also evident from the computation of $\l_t=\infty$ from the analytical computation using transfer matrix. For $\theta=\pi/(4j-1)$, the localization length decreases as $j$ increases, consistent with probability and entropic distribution observations. 
\end{itemize}
\end{shaded}
The remainder of the paper is organized as follows. In section \ref{prob}, we discuss the probability distributions for the SDC and SIC walk. Along with this we also demonstrate the relative restrictions on the number of points with a non-zero distribution and explore the categories of restricted walk by tuning the coin. In  section \ref{entropymeasure}, we explore localization from the entropy measures point of view. In particular we discuss the evidence of localization from Shannon Entropy (section \ref{shannonent}), Entanglement (section \ref{entent}) and finally compare the SDC and SIC walks through the POVM based measurements using Quantum Relative Entropy (section \ref{relent}). In section \ref{transmat}, we explore the localization phenomenon analytically using the transfer matrix approach. We demonstrate that the $L_{loc}$ as a function of the energy peaks at only certain values, demonstrating that the system is maximally localized. Further, we incorporate the SDC walk in the background of an SIC walk, as a perturbation to demonstrate that the walk approximately localizes at around the same points as dictated by the numerical simulation meaning that even in presence of a small disorder, the walk exhibits significant localization. We end the work with conclusions and future questions to be addressed in section \ref{concl}.

\section{Probability Density}\label{prob}

The Step Dependent Coin is given by,
\be
\mathbf{C}_t(\theta_1,\theta_2,\phi)=\left(\begin{array}{cccc}e^{-it\phi}\cos t\theta_1&0&0&-ie^{-it\phi}\sin t\theta_1\\0& e^{it\phi}\cos t\theta_2&-ie^{it\phi}\sin t\theta_2&0\\0& -ie^{it\phi}\sin t\theta_2&e^{it\phi}\cos t\theta_2&0\\-ie^{-it\phi}\sin t\theta_1&0&0&e^{-it\phi}\cos t\theta_1\end{array}\right)\,.
\ee
The Step Independent Coin operator is given by $\mathbf{C}_{SIC}=\mathbf{C}_{t=1}$. Teh shift operator is the same, given by,
\be
\mathbf{S}=\sum_{\vec{x}}\left(|0,\vec{x}+\vec{\a}_0\ra\la0,\vec{x}|+|1,\vec{x}+\vec{\a}_1\ra\la1,\vec{x}|+|2,\vec{x}-\vec{\a}_1\ra\la2,\vec{x}|+|3,\vec{x}-\vec{\a}_0\ra\la3,\vec{x}|\right)\,,
\ee
where $\a_0=(1,1)$ and $\a_1=(1,-1)$ and the state $|i,\vec{x}\ra=|i\ra\otimes|\vec{x}\ra$ where $|i\ra\in\mathcal{H}_\mathbf{C}$ and $|\vec{x}=(m,n)\ra\in\mathcal{H}_\mathbf{P}$. The evolution operator $\mathbf{U}=\mathbf{S}\cdot(\mathbf{C}_t\otimes\mathbf{I}_\mathbf{P})$ acts on the initial state $|\Psi_w(0)\ra$ to evolve a wave function $|\Psi_w(t)\ra$ after $t$ steps of evolution, given by,
\be
|\Psi_w(t)\ra=U^t|\Psi_w(0)\ra\,.
\ee
The final state has a general form,
\be
|\Psi_w(t)\ra=\sum_{i,\vec{x}}A^{(i)}_{\vec{x}}(t)|i,\vec{x}\ra\,.
\ee
where $|i,\vec{x}\ra=|i\ra\otimes |\vec{x}\ra\in \mathcal{H}_\mathbf{C}\otimes\mathcal{H}_\mathbf{P}$. The probability distribution, as a function of the $2d$ grid is given by,
\be
P(t,\vec{x})=\sum_{i=0}^3 |A^{(i)}_{\vec{x}}(t)|^2\,.
\ee
\begin{figure}
\begin{tabular}{cc}
\centering
  \includegraphics[width=8cm]{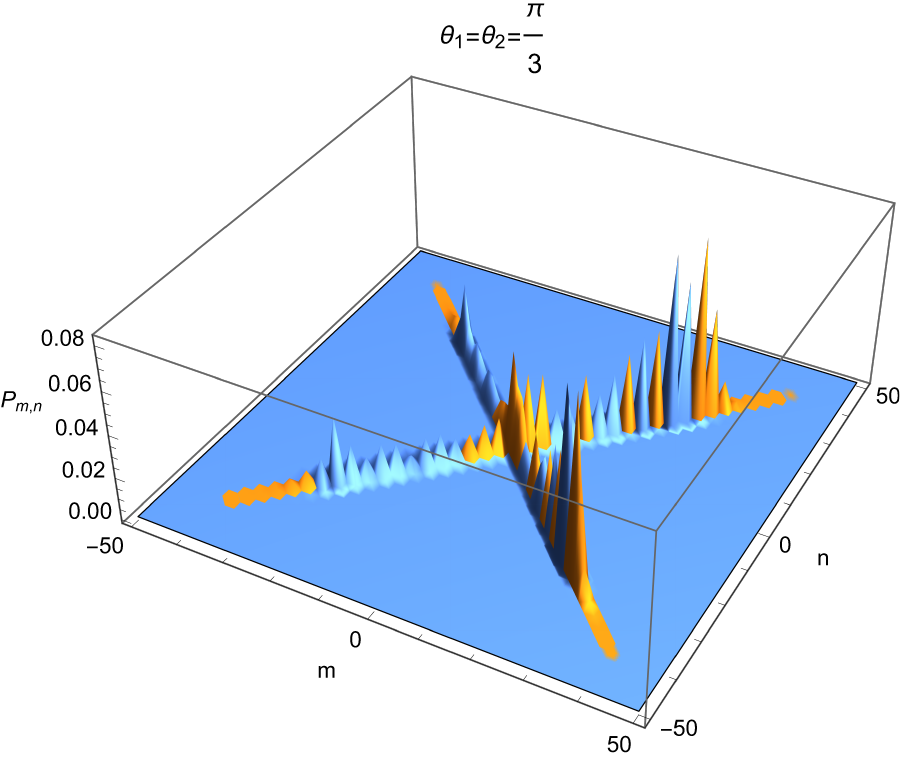} &   \includegraphics[width=8cm]{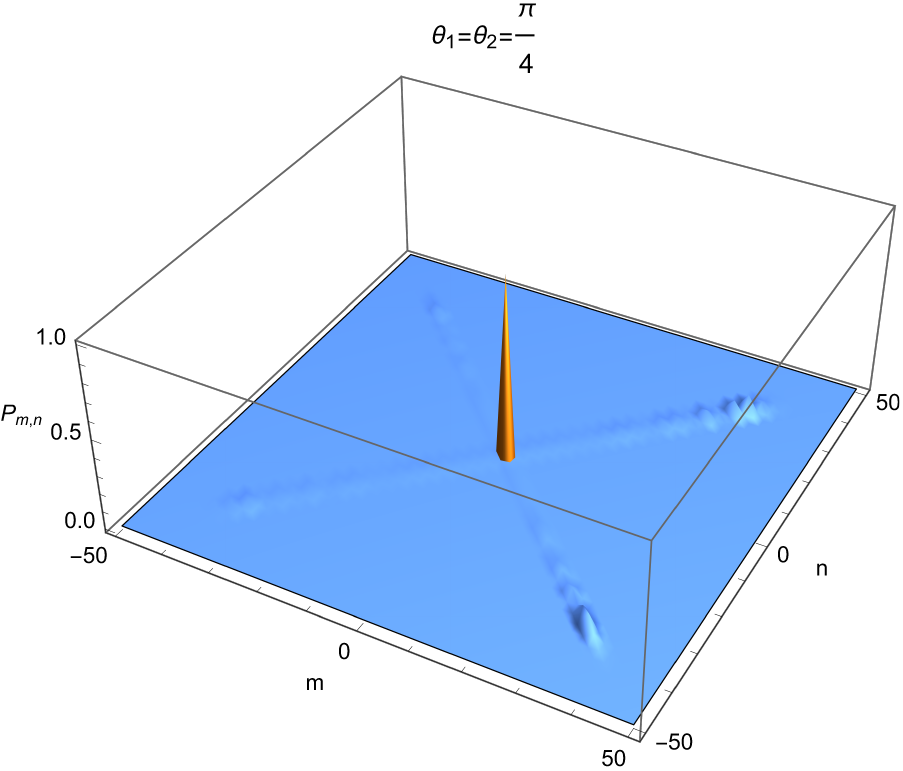} \\
(a) & (b) \\
\includegraphics[width=8cm]{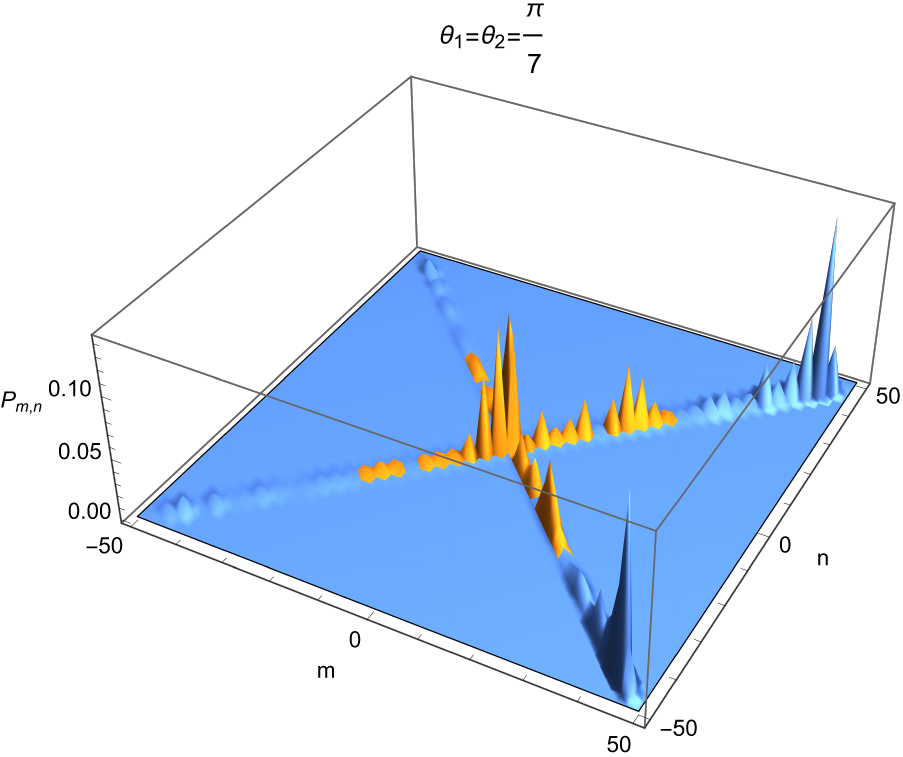} &   \includegraphics[width=8cm]{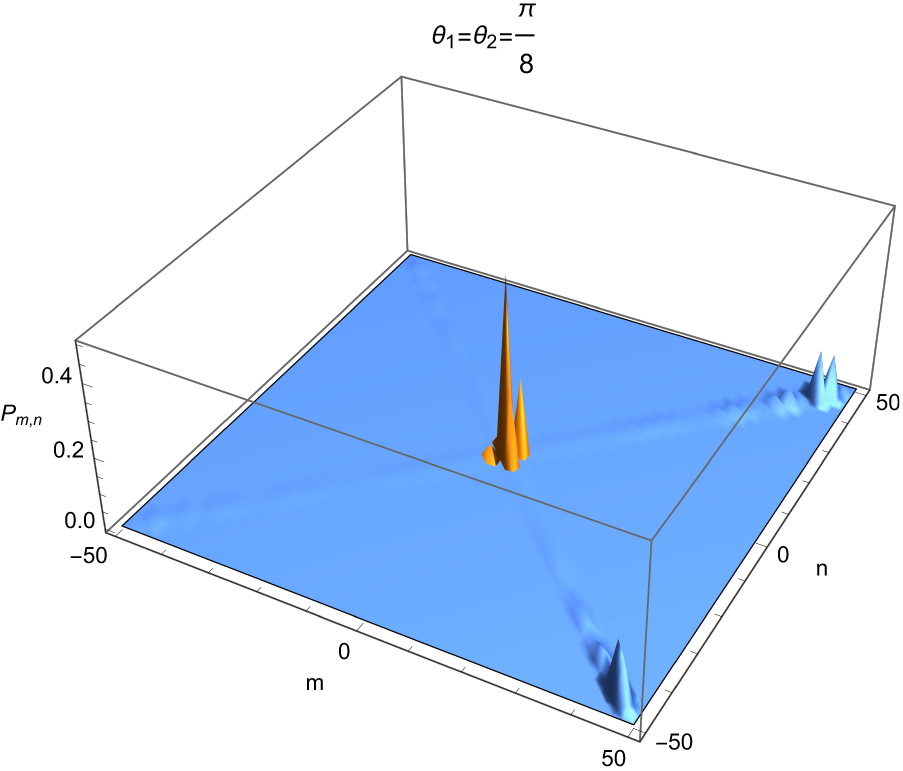} \\
(c) & (d) \\
\includegraphics[width=8cm]{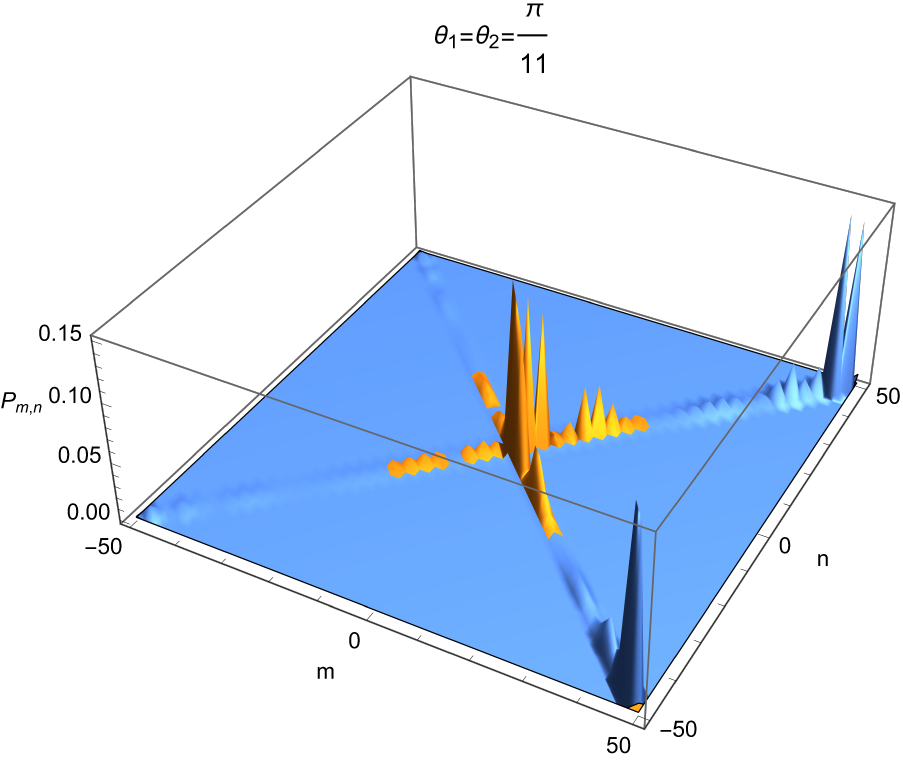} &   \includegraphics[width=8cm]{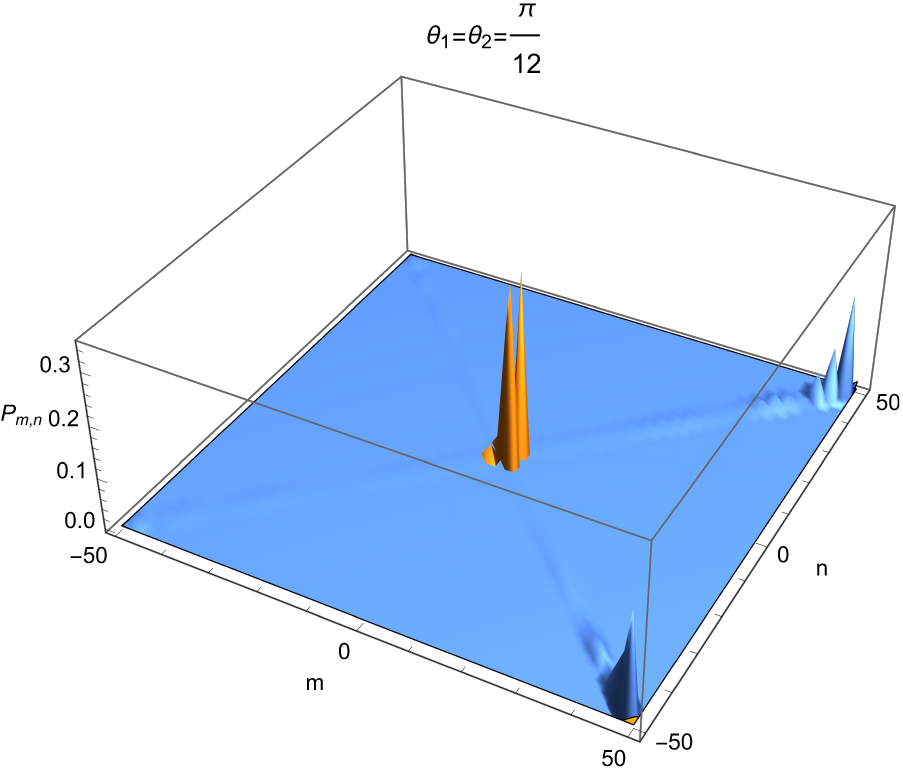} \\
(e) & (f) \\
\end{tabular}
\caption{Probability distributions for the SDC (orange peaks) and SIC (blue peaks) walks for $\theta_1=\theta_2=\theta=\pi/(4j),\pi/(4j-1)$ for $j=1,2,3$. The SDC walks are localized at the origin for $\theta=\pi/(4j)$. For $\theta=\pi/(4j-1)$, the walks start as diffused and localizes as $j$ increases. }
\label{probdis0}
\end{figure}

\begin{figure}
\begin{tabular}{cc}
\centering
  \includegraphics[width=10cm]{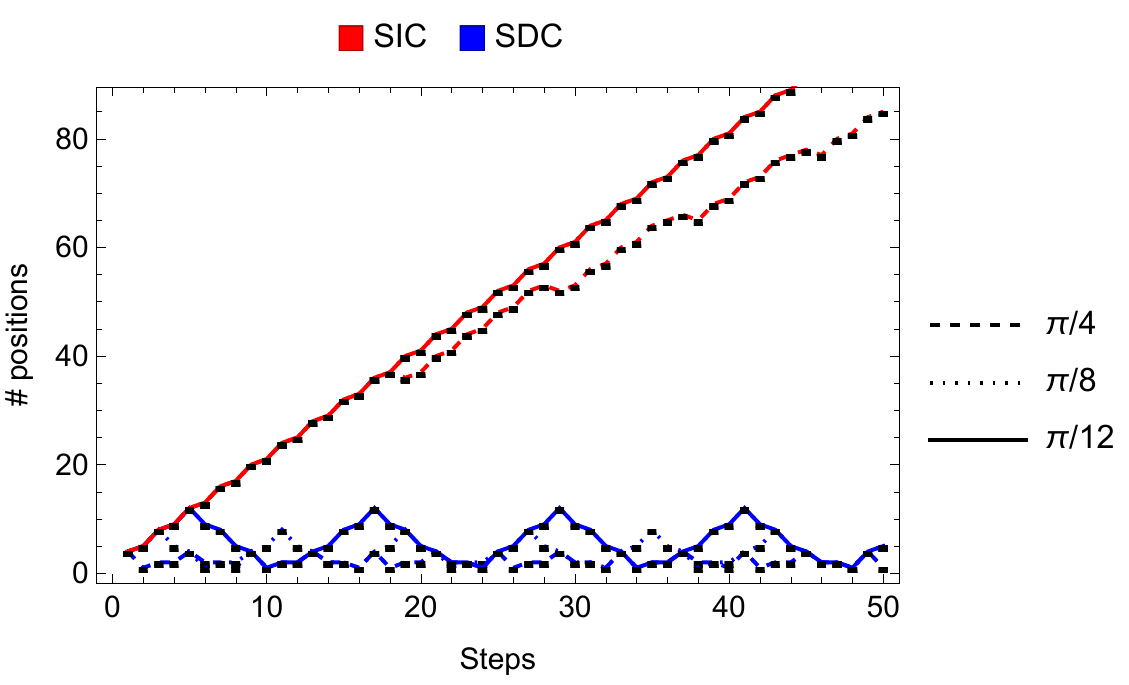} &   \includegraphics[width=10cm]{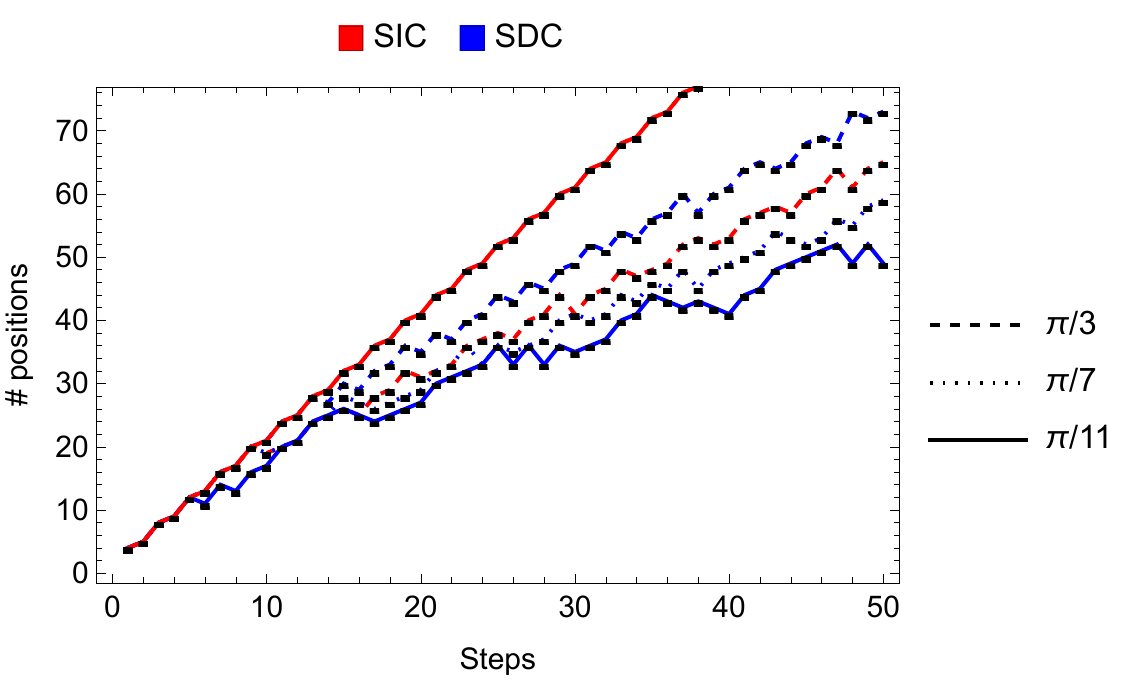} \\
(a) & (b) 
\end{tabular}
\caption{Number of positions with non-zero probability for SIC and SDC, for angles (a) $\theta_1=\theta_2=(\pi/4,\pi/8,\pi/12)$ and (b) $\theta_1=\theta_2=(\pi/3,\pi/7,\pi/11)$.  }
\label{posdis}
\end{figure}

\begin{figure}
\begin{tabular}{cc}
\centering
  \includegraphics[width=9.5cm]{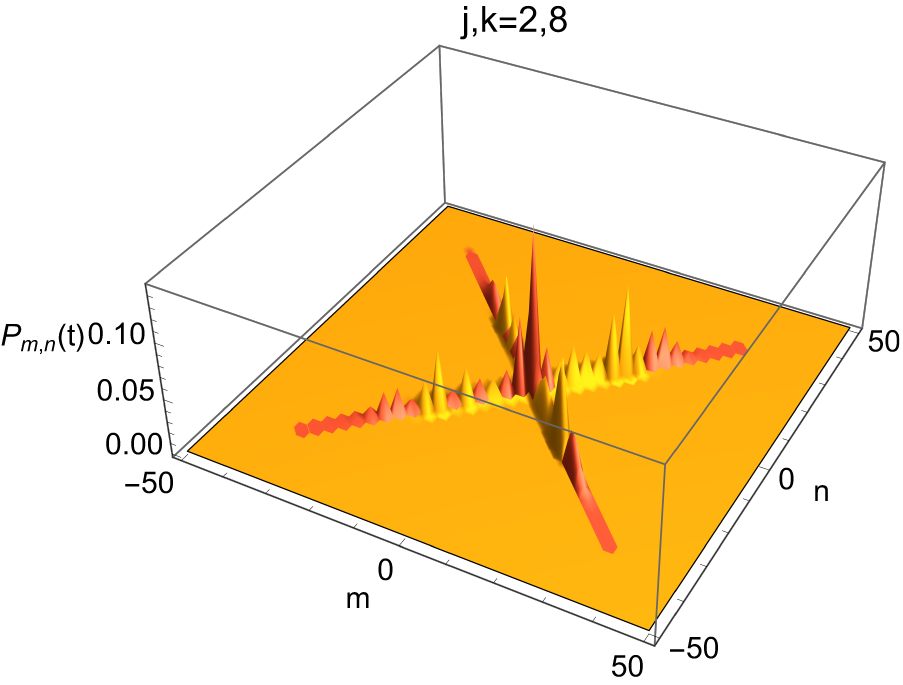} &   \includegraphics[width=9.5cm]{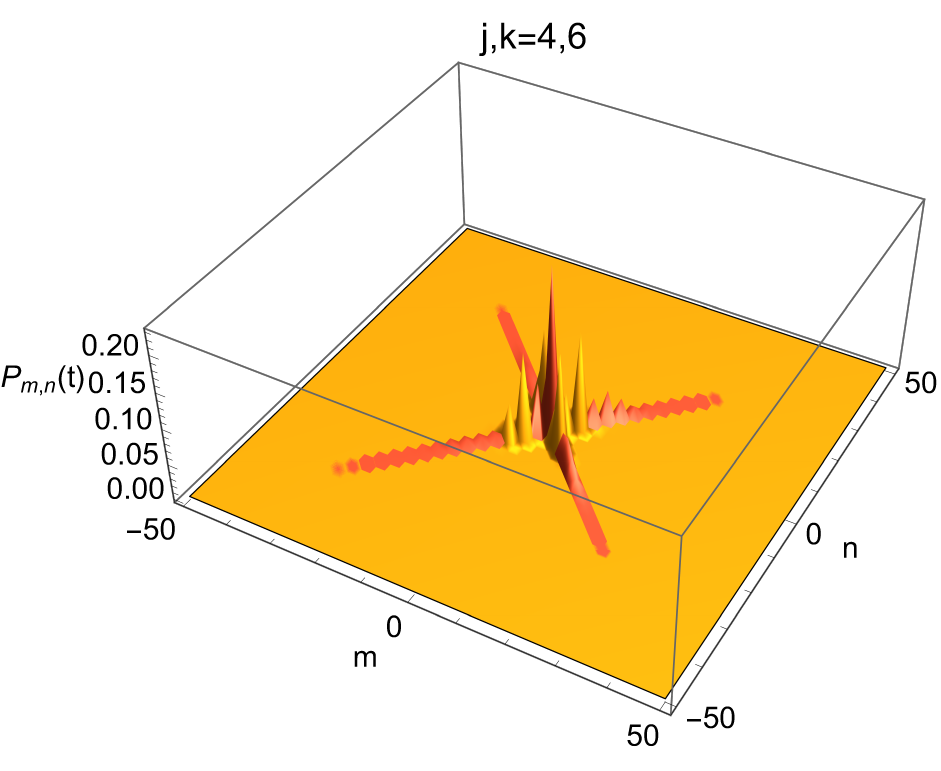} \\
(a) & (b) \\
\multicolumn{2}{c}{\includegraphics[width=9.5cm]{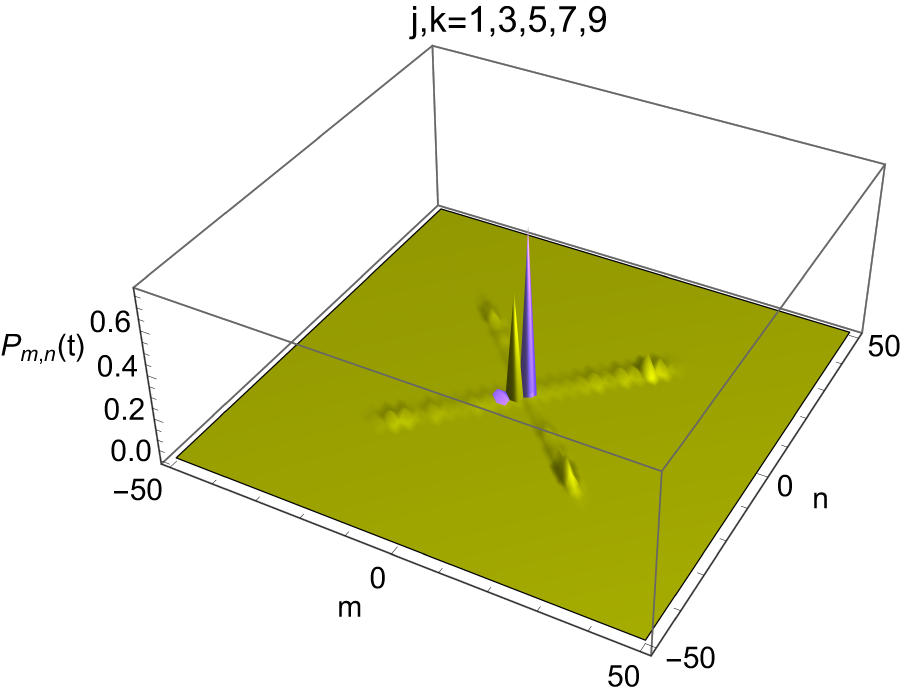} }\\
\multicolumn{2}{c}{(c) }
\end{tabular}
\caption{The corresponding SDC (yellow peaks) and SIC (red peaks) walks for (a) $j,k=2,8$ and (b) $j,k=4,6$ and (c) $j,k=1,3,5,7,9$. (a) demonstrates a semi-classical/quantum-like multi-modal walk.(b) demonstrates a compact classical-like walk with nearly symmetric distributions around the origin. (c) demonstrates a completely localized walk. }
\label{probdis}
\end{figure}
\begin{figure}
\begin{tabular}{cc}
\centering
  \includegraphics[width=10cm]{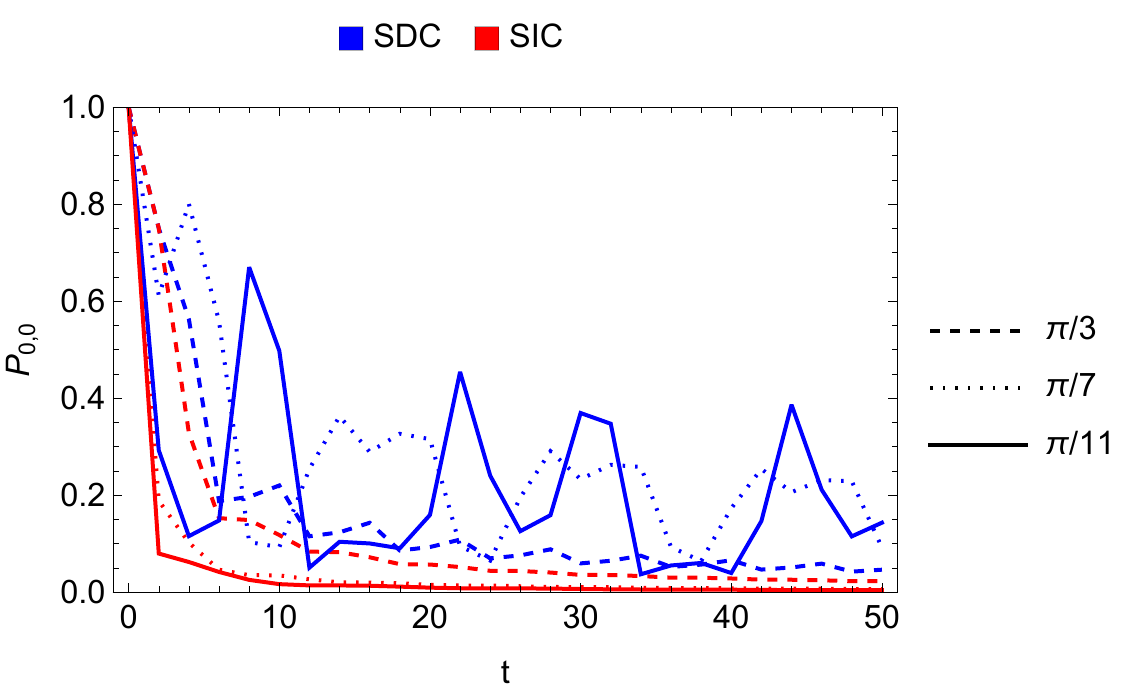} &   \includegraphics[width=10cm]{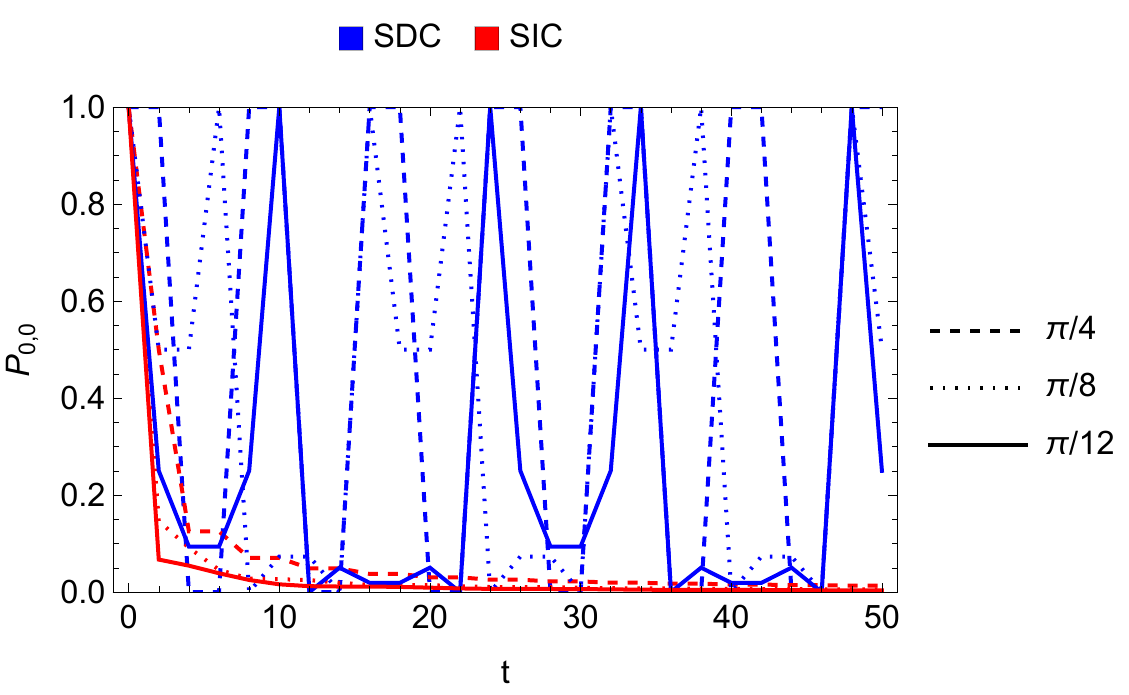} \\
(a) & (b) 
\end{tabular}
\caption{Return probability for (a) $\theta_1=\theta_2=(\pi/3,\pi/7,\pi/11)$ and (b) $\theta_1=\theta_2=(\pi/4,\pi/8,\pi/12)$  }
\label{retprob}
\end{figure}

We demonstrate the probability distributions for the SDC and SIC walks for $\theta_1=\theta_2=\theta$ and $\phi=0$. In figure \ref{probdis0}, we demonstrate the SDC and SIC walks for $\theta\pi/3\,,\pi/4\,, \pi/7\,,\pi/8\,, \pi/11$ and $\pi/12$. For even values of the denominator, we find that the SDC walk is localized at and around the origin while the SIC walk spreads out. For odd values of the denominator, the SDC and SIC walks start out with significant spreads. However as $\theta$ decreases, the SDC walk migrates to the origin. The same feature is exhibited in figure \ref{posdis}, where we compute the number of positions with non-zero probability distribution and compare for the SDC and SIC walks. We again find that for even values of the denominator, the SDC walk is an oscillatory and bounded function of the time steps while the SIC walk linearly grows with time. For odd values of the denominator in $\theta$, the SDC and SIC walks start off with a linear growth with time. As $\theta$ decreases, the plot for the SDC walk comes down indicating that the number of positions with non-zero probability distribution decreases signifying that the wave-function becomes localized. 
In figure \ref{probdis}, we consider tuning the coin parameters to categorize the quantum walks. To begin with, we start with $\theta_1=\theta_2=\pi/3$ and $\phi=0$, where the walk is quantum in nature and start tuning $\theta=\pi/3(1+j/10)$ for $1\leq j\leq 10$. For $j=1,3,5,7,9$, the transition is to a {\it fully localized walk}. For $j=2,8$, it is a {\it compact semi-classical/quantum walk}. For $j=4,6$, the walk is a {\it compact classical} walk.  This illustrates that the quantum regime can be tuned to a classical and semi-classical compact or a fully localized walk with the step dependent coin. Finally we explore the return probability for the wave function to the origin in figure \ref{retprob},  given by $P_{0,0}(t)$ as a function of the time step $t$. For SIC walk, the wave function spreads out and after $t$ steps, the probability of the wave function at the origin decreases exponentially. For the SDC walk however, we see that for $\theta=\pi/(4j-1)$ the return probability follows the SIC distribution with certain peaks. However as $j$ increases, the peaks become significant implying that the walk accumulates at the origin. For $\theta=\pi/(4j)$, we see periodic peaks $P_{0,0}(t)=1$ in the return probability meaning that the walk is completely localized at the origin. At the steps, where the $P_{0,0}(t)=0$, implies that the walk is localized at some other point as exhibited in figure \ref{probdis0}. 

\section{Entropic Measures}\label{entropymeasure}
In addition to the kinematical features exhibited by the step dependent coin in the random walk, we would examine the dynamical features of the walk as well from entropic measurements. For our purpose, we would choose to demonstrate the localization of the walk through the measures of (a) Shannon Entropy, (b) Entanglement and (c) Quantum Relative Entropy between the density matrix operators. Please note that the measurements will also depend on the initial state. For our purpose, we choose a tensor product initial state so that the walk becomes entangled solely due to the coin operator. Our initial state is,
\be
|\Psi_w(0)\ra=\frac{1}{\sqrt{2}}\left(\begin{array}{c}1\\i\\0\\0\end{array}\right)\otimes|0,0\ra\,.
\ee
Note that this choice of initial state is by no means unique and any tensor product state can be chosen. However, the basic aspects of localization will remain unaltered and we choose to demonstrate those features through a simple choice.

\subsection{Shannon Entropy}\label{shannonent}
\begin{figure}
\begin{tabular}{cc}
\centering
  \includegraphics[width=9.5cm]{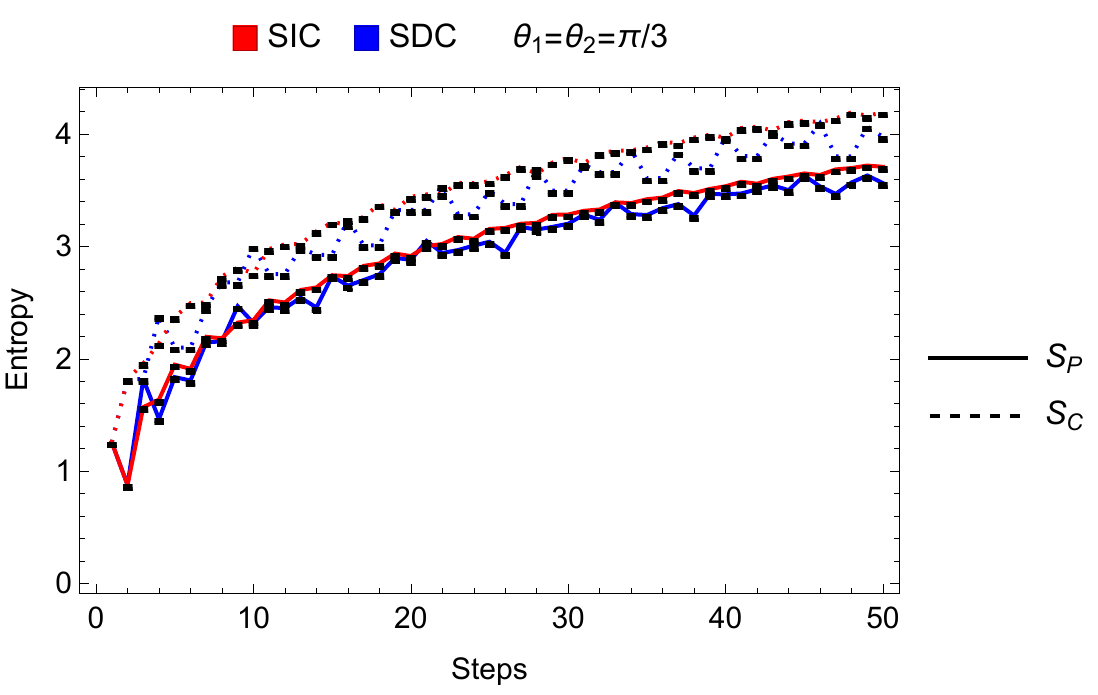} &   \includegraphics[width=9.5cm]{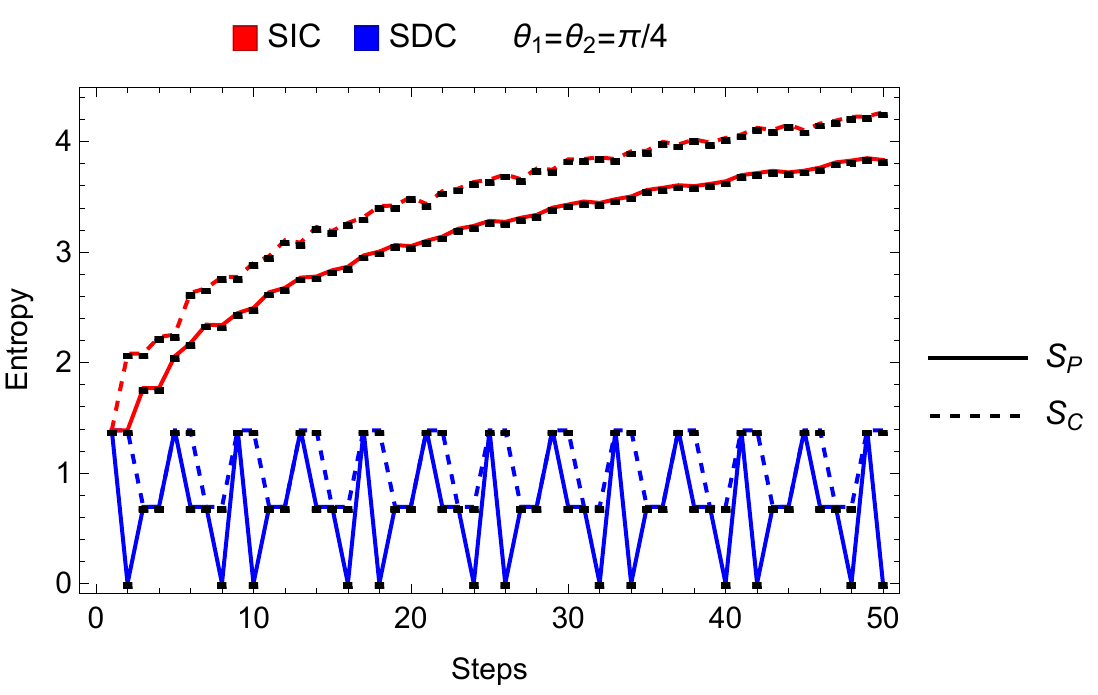}\\
  (a) & (b)\\
  \includegraphics[width=9.5cm]{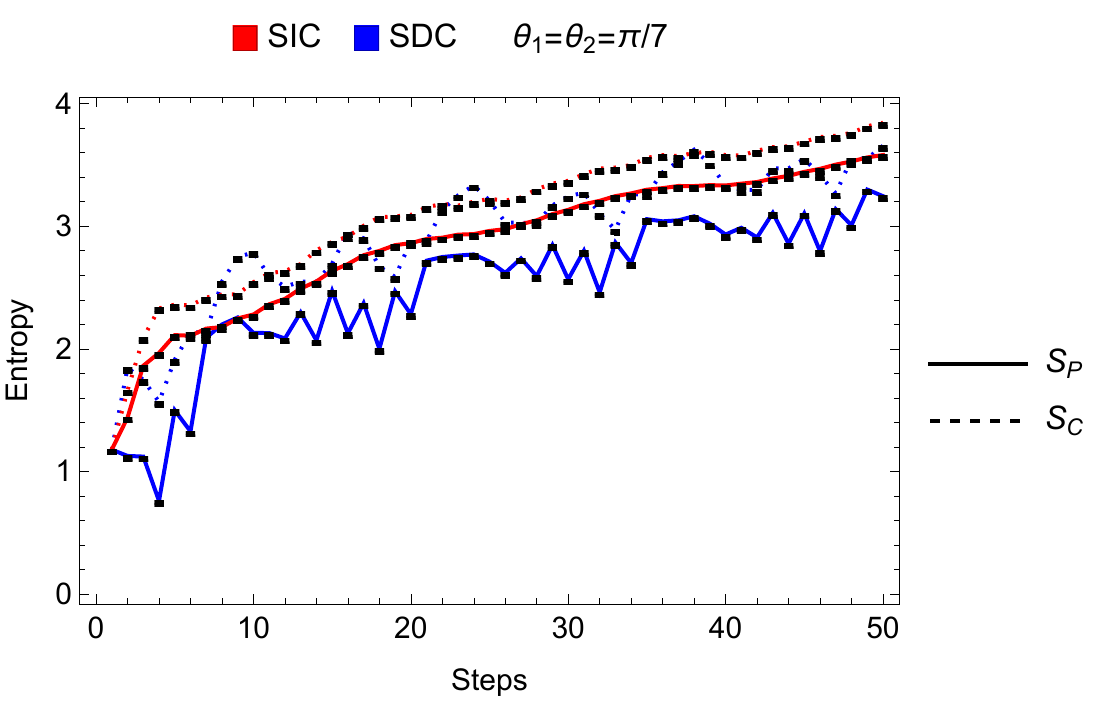} &   \includegraphics[width=9.5cm]{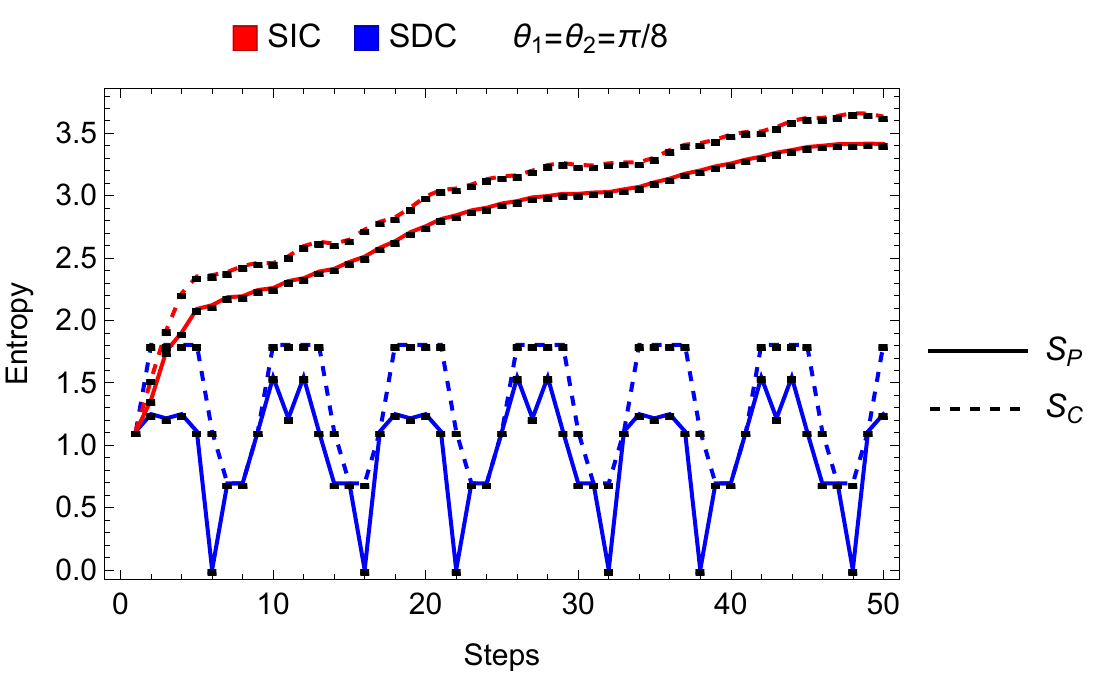} \\
(c) & (d) \\
\includegraphics[width=9.5cm]{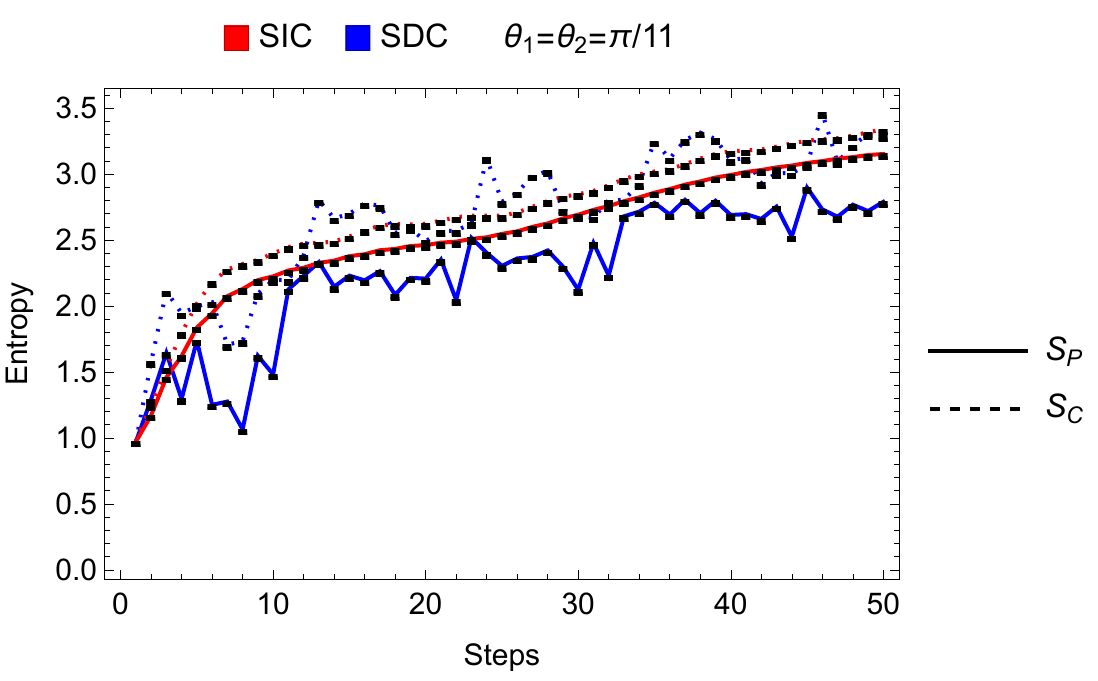} &   \includegraphics[width=9.5cm]{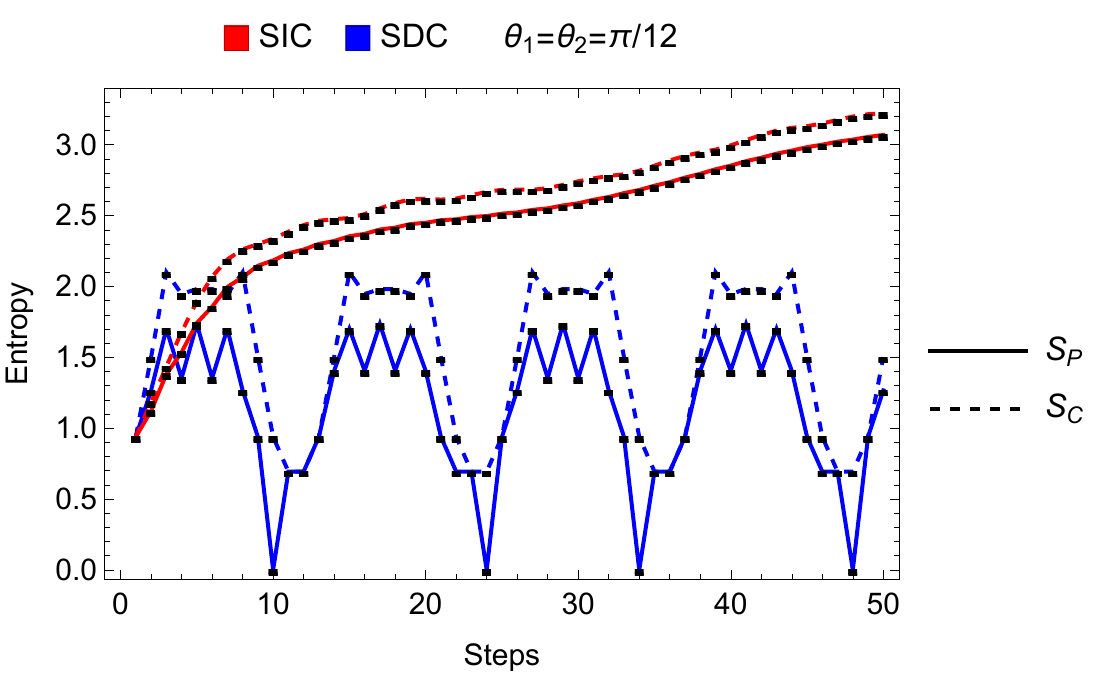} \\
(e) & (f) 
\end{tabular}
\caption{Shannon Entropy for position space $(S_P)$ and coin space $(S_C)$ for (a) $\theta_1=\theta_2=\pi/3$, (b) $\theta_1=\theta_2=\pi/4$, (c) $\theta_1=\theta_2=\pi/7$, (d) $\theta_1=\theta_2=\pi/8$, (e) $\theta_1=\theta_2=\pi/11$ and (f) $\theta_1=\theta_2=\pi/12$ for the SDC and SIC walks. SDC walks are much more suppressed than the SIC walks. }
\label{shannondis}
\end{figure}

We will compute and compare the Shannon Entropy for the position space and coin space for SDC and SIC walks \cite{Panahiyan_2018}. These measures are defined as,
\be
S_P=-\sum_{m,n=-t}^t P_{m,n}(t)\log P_{m,n}(t)\,, \ S_C=-\sum_{i=0}^3 P_i(t)\log P_i(t)\,. 
\ee
where,
\be
P_{m,n}(t)=\sum_{i=0}^3 |A^{(i)}_{m,n}(t)|^2\,,\ \ P_i(t)=\sum_{m,n=-t}^t |A^{(i)}_{m,n}(t)|^2\,.
\ee
Also note that as the wave function localizes in a particular point $(m,n)$, $P_{m,n}(t)=1$ and zero elsewhere. Consequently, $S_P=0$. For the $S_C$, the Shannon entropy however oscillates between and maxima and minima (even for a completely localized state). We compare the Shannon entropy for a range of $\theta$ values of the coin parameters in figure \ref{shannondis}. Completely localized states are obtained for $\theta_1=\theta_2=\theta=\pi/(4j)$ in sharp contrast to its SIC counterpart with same $\theta$. This is exhibited by a vanishing Shannon Entropy due to a separable state. For $\theta_1=\theta_2=\theta=\pi/(4j-1)$ however, the walk is not completely localized. For $j=1$, the SDC and SIC walk has maximal overlap. However as $j$ increases, the walk becomes more localized and $S_P$ decreases in contrast to its SIC counterpart. In these cases, however the $S_C$ increases over the SIC counterpart hinting at the growing localization of the walk. 

\subsection{Entanglement}\label{entent}
\begin{figure}
\begin{subfigure}{.5\textwidth}
  \centering
  \includegraphics[width=\linewidth]{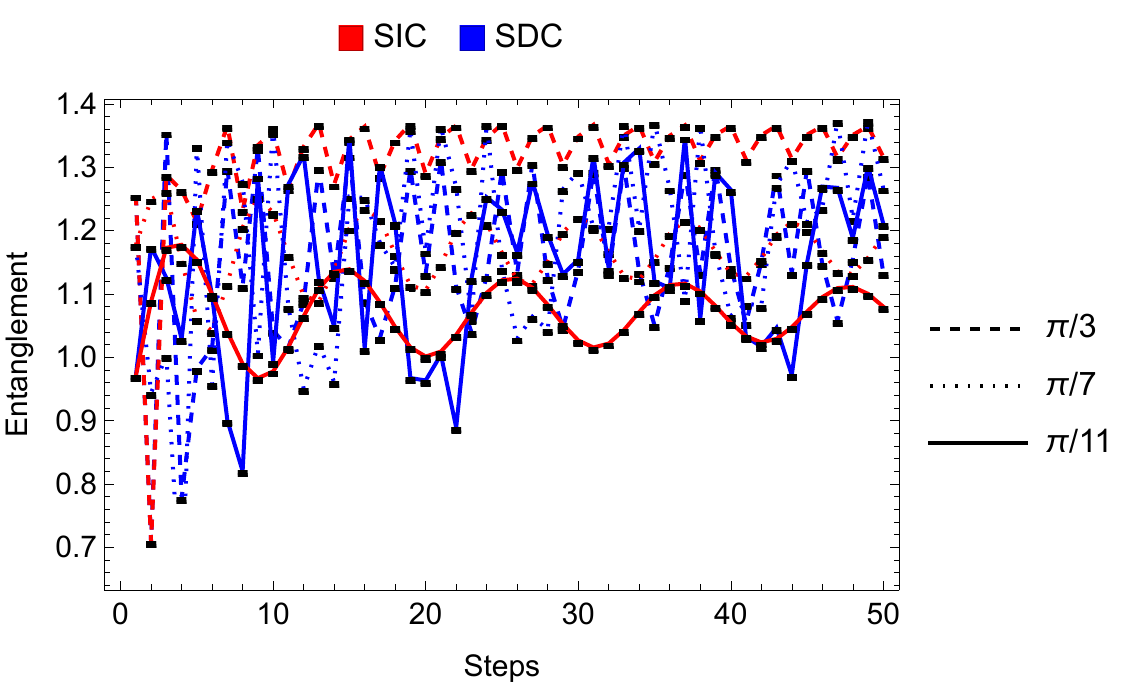}
  \caption{}
  \label{fig:sfig1}
\end{subfigure}%
\begin{subfigure}{.5\textwidth}
  \centering
  \includegraphics[width=\linewidth]{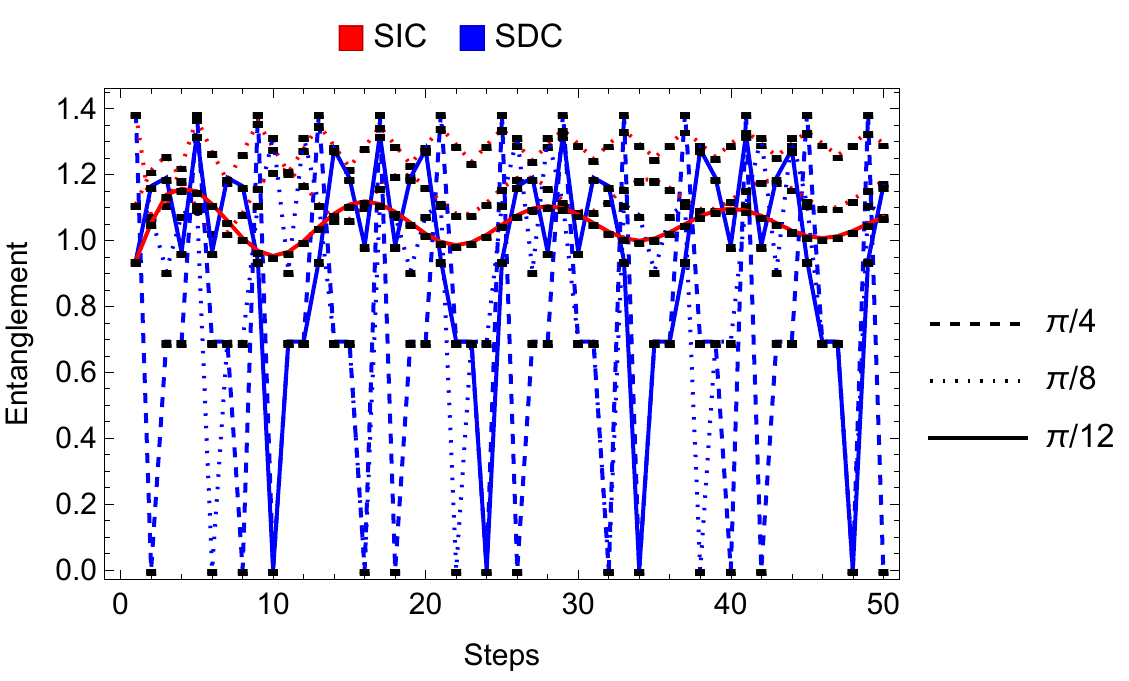}
  \caption{}
  \label{fig:sfig2}
\end{subfigure}
\caption{Entanglement Entropy for (a) $\theta_1=\theta_2=(\pi/3,\pi/7,\pi/11)$ and (b) $\theta_1=\theta_2=(\pi/4,\pi/8,\pi/12)$ for the SDC and SIC walks. }
\label{entanglement}
\end{figure}

The second entropic measure is the entanglement entropy between the position and coin (spin) degrees of freedom. The reduced density matrix after tracing over the position coordinates is given by,
\be
\widetilde{\r}(t)=\tr_{pos}\r(t)=\sum_{m,n=-t}^t\left(\begin{array}{cccc}|A^{(0)}_{m,n}|^2& A^{(0)}_{m,n}A^{(1)\star}_{m,n}&A^{(0)}_{m,n}A^{(2)\star}_{m,n}&A^{(0)}_{m,n}A^{(3)\star}_{m,n}\\|A^{(1)}_{m,n}A^{(0)\star}_{m,n}& |A^{(1)}_{m,n}|^2&A^{(1)}_{m,n}A^{(2)\star}_{m,n}&A^{(1)}_{m,n}A^{(3)\star}_{m,n}\\ A^{(2)}_{m,n}A^{(0)\star}_{m,n}&A^{(2)}_{m,n}A^{(1)\star}_{m,n}&|A^{(2)}_{m,n}|^2&A^{(2)}_{m,n}A^{(3)\star}_{m,n}\\A^{(3)}_{m,n}A^{(0)\star}_{m,n}&A^{(3)}_{m,n}A^{(1)\star}_{m,n}&A^{(3)}_{m,n}A^{(2)\star}_{m,n}&|A^{(3)}_{m,n}|^2\end{array}\right)\,.
\ee
The entanglement is given by,
\be
E(t)=-\tr\widetilde{\r}(t)\log\widetilde{\r}(t)\,.
\ee
If the final state is a completely localized state as happens for $\theta_1=\theta_2=\theta=\pi/(4j)$, then,
\be
|\Psi_w(T)\ra=\psi_c\otimes |p,q\ra\,,
\ee
is a separable state. The corresponding reduced density matrix is a pure state and the entanglement is zero. This is demonstrated in figure \ref{fig:sfig2}, where the entanglement drops to zero for certain time steps for $\theta_1=\theta_2=\pi/(4j)$. Figure \ref{fig:sfig1}, demonstrates the same for $\theta=\pi/(4j-1)$. The walks are not completely localized but the walks become more localized with increasing $j$ which is exhibited in the certain dips in the entanglement.

\subsection{Relative Entropy}\label{relent}
\begin{figure}
\begin{tabular}{cc}
\centering
  \includegraphics[width=9.5cm]{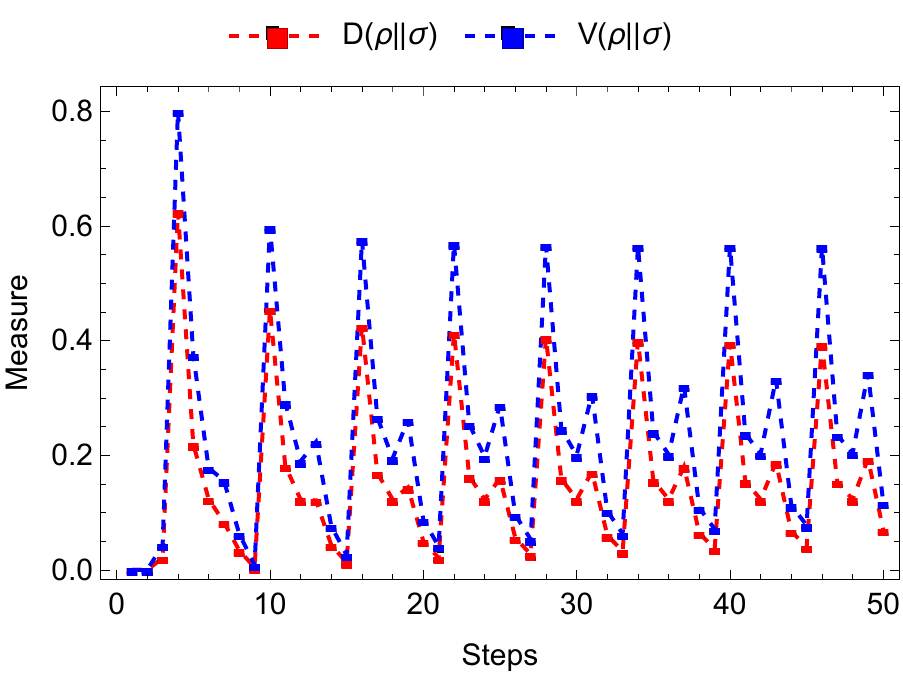} &   \includegraphics[width=9.5cm]{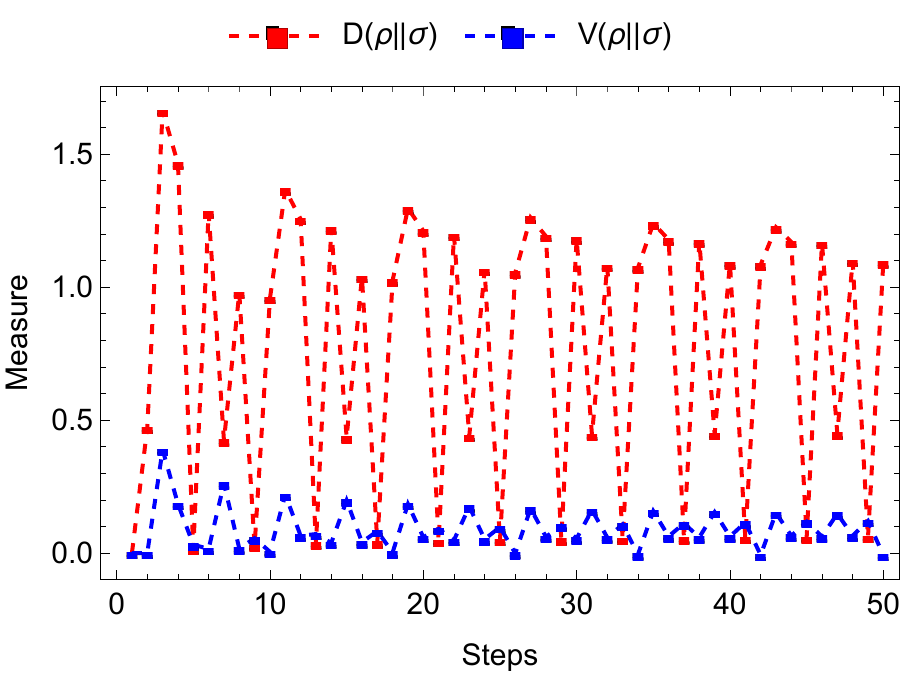} \\
(a) & (b) \\
 \includegraphics[width=9.5cm]{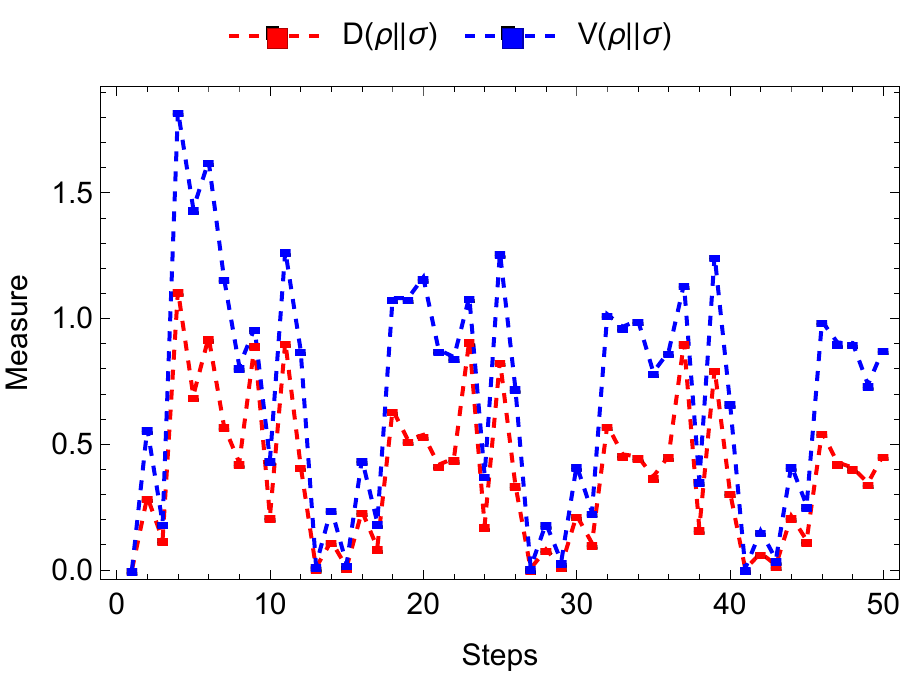} &   \includegraphics[width=9.5cm]{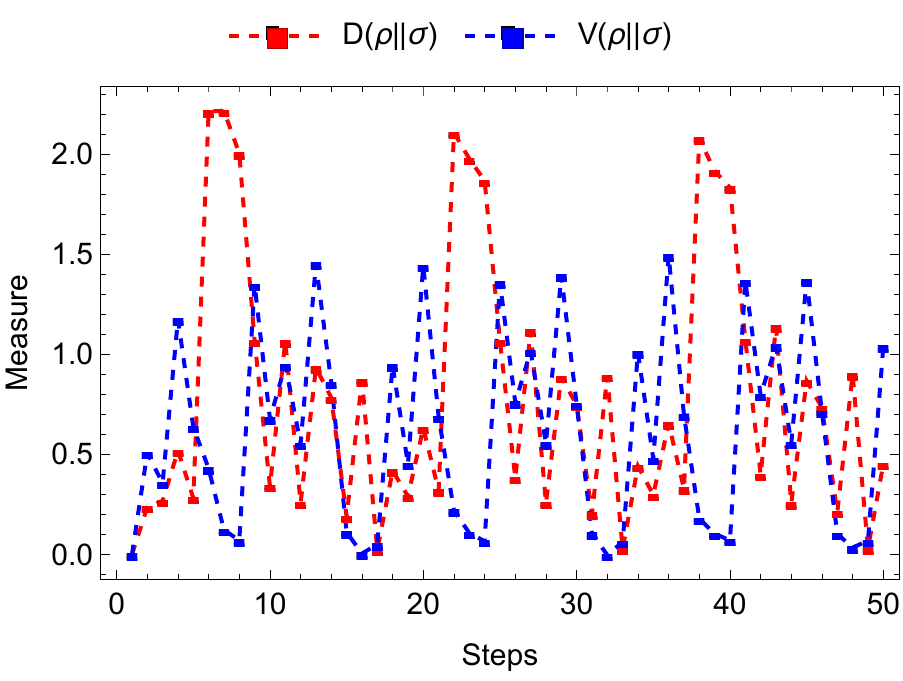} \\
(c) & (d) \\
 \includegraphics[width=9.5cm]{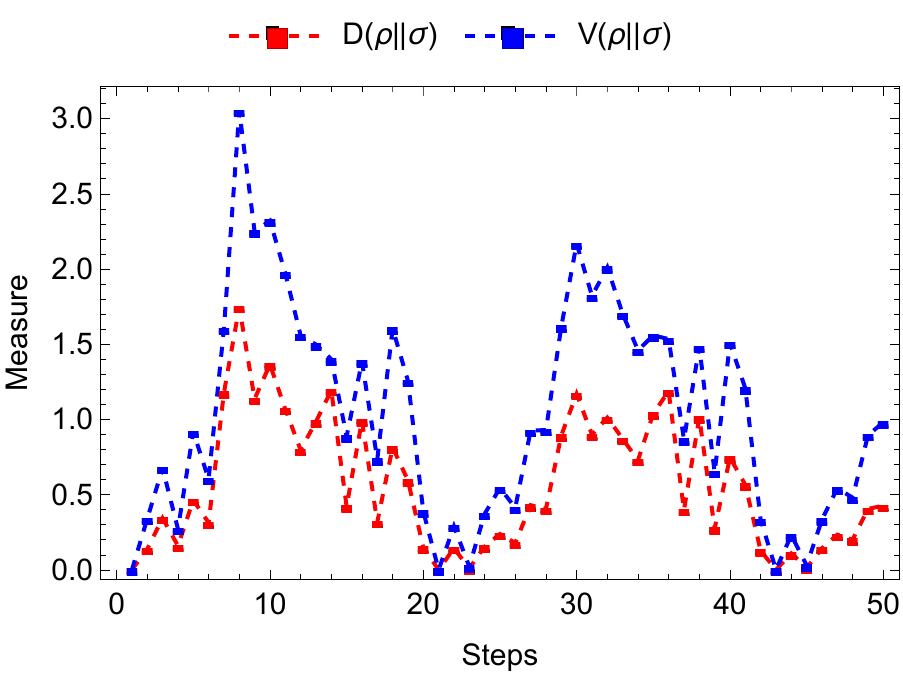} &   \includegraphics[width=9.5cm]{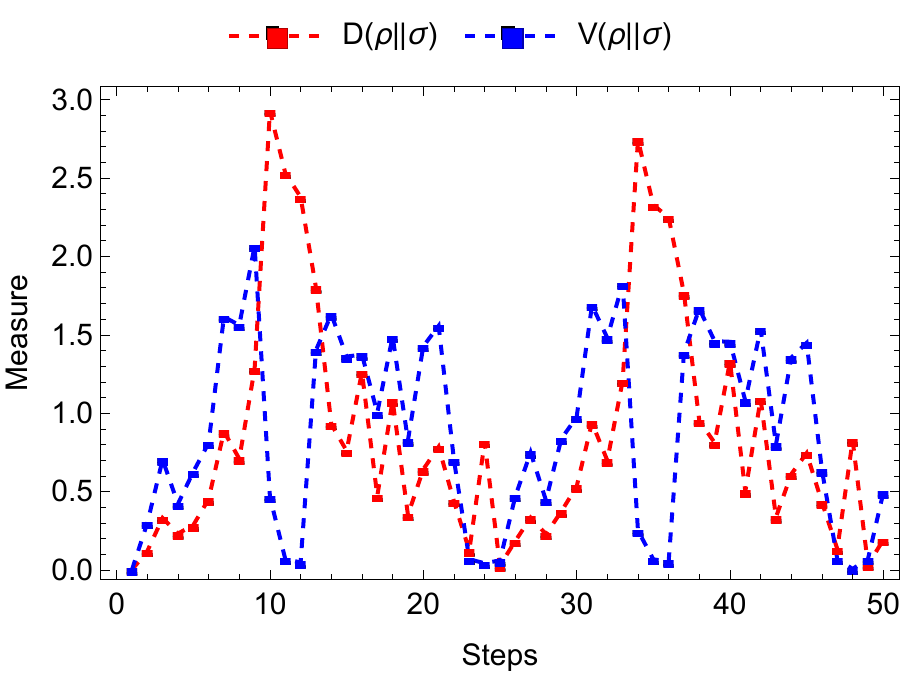} \\
(e) & (f) 
\end{tabular}
\caption{$D(\r||\s)$ and $V(\r||\s)$ for (a) $\theta_1=\theta_2=\pi/3$, (b) $\theta_1=\theta_2=\pi/4$, (c) $\theta_1=\theta_2=\pi/7$, (d) $\theta_1=\theta_2=\pi/8$, (e) $\theta_1=\theta_2=\pi/11$ and (f) $\theta_1=\theta_2=\pi/12$ for the SDC and SIC walks. }
\label{relendis}
\end{figure}

Finally the third entropic measure is a POVM between the density matrix operators for the SDC and SIC walks. We consider Quantum Relative Entropy (QRE) and Quantum Information Variance (QIV) between the density matrices for SDC and SIC walks. The QRE is defined by \cite{Leditzky,Leditzky_2016,Leditzky_20161},
\be
D(\r||\s)=\tr\r(\log\r-\log\s)\,,
\ee
and the QIV is given by,
\be
V(\r||\s)=\tr\r(\log\r-\log\s)^2-D(\r||\s)^2\,.
\ee
The operators $\r$ and $\s$ are two positive valued reduced density matrices $\r$ and $\s$ for the step dependent and step independent coin operator. For the walks, we have chosen, $\theta_1=\theta_2=\theta$ and $\phi=0$ as previously. We further group the plots in terms of $\theta=\pi/(4j-1)$ and $\theta=\pi/(4j)$ for $j=1,2,3$ as demonstrated in figure \ref{relendis}. Note that both $D(\r||\s)$ and $V(\r||\s)$ increase as $j$ increases (both for odd and even values of the denominator). For odd values of the denominator $V(\r||\s)\geq D(\r||\s)$ while for even values, $D(\r||\s)\geq V(\r||\s)$. This observation is also consistent with the probability distributions for the chosen values of $\theta$. For odd values of the denominator,  both the SDC and SIC walks have significant overlaps and the difference between the SDC and SIC becomes more pronounced as $j$ increases. The exact opposite happens for even values of the denominator (as $j$ increases). High values of relative entropy and low values of QIV signifies a pronounced difference between the SDC and SIC walks.  For $V(\r||\s)=0$, 
\be\label{zerovar}
\tr\r(\log\r-\log\s)^2=\left(\tr\r(\log\r-\log\s)\right)^2\,.
\ee
We start by writing,
\be
\r(t)=\l_i|\Phi_i\ra\la\Phi_i|\,, \ \ \s(t)=\l'_i|\Phi'_i\ra\la\Phi'_i|\,,
\ee
in terms of the eigen basis. Thus, using $c_{ik}=\la\Phi_i|\Phi'_k\ra$,
\be
\r(\log \r-\log\s)^2=\l_i(\log\l_i)^2|\Phi_i\ra\la\Phi_i|+\l_i(\log\l'_j)^2 |\Phi'_j\ra\la\Phi'_j|-2\l_i\log\l_i\log\l'_kc_{ik}|\Phi_i\ra\la\Phi'_k|\,.
\ee
Then,
\be
\tr\r(\log\r-\log\s)^2=\l_i(\log\l_i)^2+\l_i(\log\l'_j)^2|c_{ij}|^2-2\l_i \log\l_i\log\l'_j|c_{ij}|^2\,.
\ee
while,
\be
D(\r||\s)=\l_i\log\l_i-\l_i\log\l'_j|c_{ij}|^2\,.
\ee
The zeroes of $D(\r||\s)$ imply that $\r=\s$ or precisely that the SDC and SIC walks coincide. $V(\r||\s)=0$ with $D(\r||\s)\neq0$, implies that the final states for both SDC and SIC walks come arbitrarily close to a maximally entangled state.

\section{Transfer Matrix}\label{transmat}
In this section we formulate the transfer matrix for the localization problem. We will go the reverse way of constructing the Hamiltonian from the transfer matrix. Given a wave function, $\Psi_{\vec{x}}=(A^{(0)}_{\vec{x}},A^{(1)}_{\vec{x}-\vec{\b}},A^{(2)}_{\vec{x}},A^{(3)}_{\vec{x}-\vec{\a}})$, for $\vec{x}=(m,n)$, $\a=(1,1)$ and $\b=(1,-1)$, we can write, 
\be
\left(\begin{array}{c}A^{(0)}_{\vec{x}+2\vec{\a}}\\ A^{(1)}_{\vec{x}+\vec{\b}}\\A^{(2)}_{\vec{x}+2\vec{\b}}\\A^{(3)}_{\vec{x}+\vec{\a}}\end{array}\right)=\mathbf{T}\left(\begin{array}{c}A^{(0)}_{\vec{x}}\\ A^{(1)}_{\vec{x}}\\A^{(2)}_{\vec{x}}\\A^{(3)}_{\vec{x}}\end{array}\right)\,.
\ee
The form of the transfer matrix for our $2d$ walk is given by,
\be
\mathbf{T}=\left(\begin{array}{cccc}e^{-it\phi+i\w_1}\sec t\theta_1&0&0& -i \tan t\theta_1\\ 0&e^{i t\phi+i\w_2}\sec t\theta_2&-i\tan t\theta_2&0\\ 0&i\tan t\theta_2& e^{-it\phi-i\w_2}\sec t\theta_2&0\\ i\tan t\theta_1&0&0&e^{it\phi-i\w_1}\sec t\theta_1\end{array}\right)\,.
\ee
$\w_1\,, \w_2$ are the eigenvalues of the spectrum satisfying the dispersion relation,
\be
\cos\w_1=\cos\theta_1\cos(k_1+k_2)\,, \cos\w_2=\cos\theta_2\cos(k_1-k_2)\,.
\ee
We define the Lyapunov exponent by,
\be
\l_t=\frac{1}{t}\sum_{\vec{x}=1}^t \log|\Psi_{\vec{x}}|\,,
\ee
with an initial condition $\Psi_{\vec{x}}^{(0)}=(1,0,0,0)$. The localization length is defined by the inverse of the Lyapunov exponent,
\be
L_{loc}=\frac{1}{\l_t}\,.
\ee
Note that for $\theta=\pi/(4j)$, we note that the localization length is zero which is consistent with the probability distribution demonstrating that the wave function is completely localized at the position of the initial wave function. For $\theta=\pi/(4j-1)$, $j=1,2,3$, we plot the localization length $(L_{loc})$ as a function of the frequency $(\w)$,

\begin{figure}
\begin{tabular}{cc}
\centering
  \includegraphics[width=10cm]{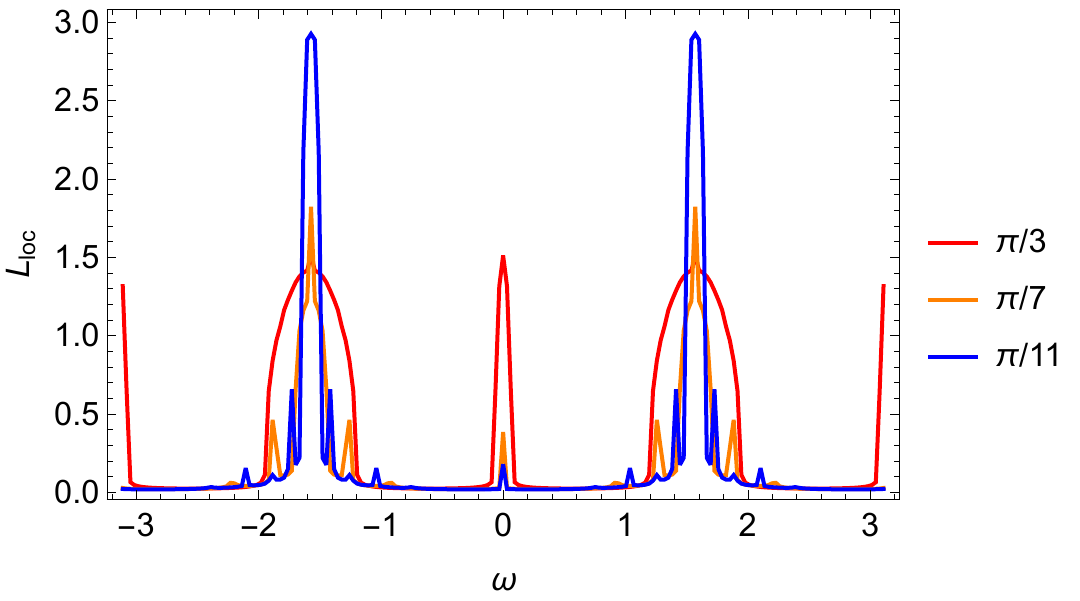} &   \includegraphics[width=10cm]{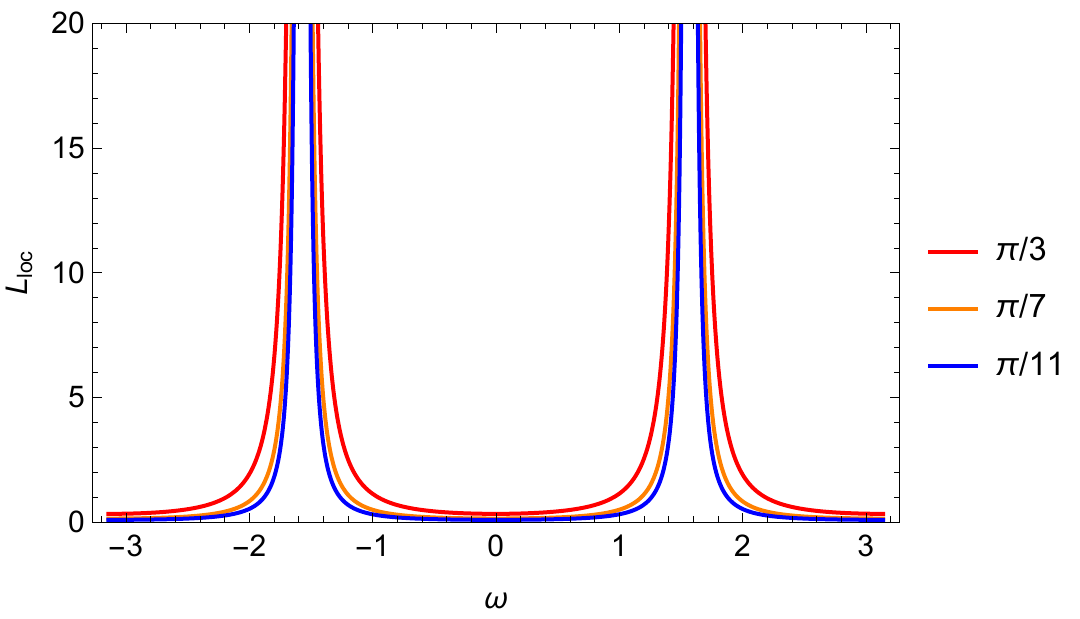} \\
(a) & (b) 
\end{tabular}
\caption{Localization length as a function of the frequency for $\theta=\pi/3,\pi/7$ and $\pi/11$.  The localization length peaks around $\w\sim\pm \pi/2$. (a) is the exact computation while (b) is the approximate analytic computation.}
\label{localdis}
\end{figure}
Having determined the localization properties of the step dependent coin, we will now use this step dependent coin as a perturbation to an otherwise SIC $2d$ random walk.  The eigenvalues are,
\be
\l=e^{\pm i\bar{\w}_1}\,, e^{\pm i\bar{\w}_2}\,,
\ee
where, $\cos\bar{\w}_1=\cos(\w_1)\sec(\theta_1)$ and $\cos\bar{\w}_2=\cos(\w_2)\sec(\theta_2)$. 
We can combine with a similarity transformation to bring the transfer matrix to a diagonal form. This similarity transform is,
\be
S=\left(\begin{array}{cccc}\tan \theta_1&0&0&-\xi_1\\0&\tan \theta_2&-\xi_2&0\\0&\xi_2&-\tan \theta_2\\ \xi_1&0&0&-\tan \theta_1\end{array}\right)\,.
\ee
where $\xi_1=\sin(\w_1)\sec\theta_1-\sin\bar{\w}_1$, $\xi_2=\sin(\w_2)\sec\theta_2-\sin\bar{\w}_2$. The diagonalized matrix is given by,
\be
S^{-1}T_{SIC}S=\left(\begin{array}{cccc}e^{i\bar{\w}_1}&0&0&0\\0&e^{i\bar{\w}_2}&0&0\\0&0&e^{-i\bar{\w}_2}&0\\0&0&0&e^{-i\bar{\w}_1}\end{array}\right)\,.
\ee
In the same basis, the SDC which will be imposed as a disorder perturbation, will yield,
\be
S^{-1}T_{SDC}S=\left(\begin{array}{cccc}\a(t)&0&0&\b(t)\\ 0&\g(t)&\d(t)&0\\0&\d(t)^\star&\g(t)^\star&0\\ \b(t)^\star&0&0&\a(t)^\star\end{array}\right)\,.
\ee
where,
\begin{align}
\begin{split}
\a(t)&=\frac{\xi _1^2 e^{-i \omega _1} \sec \left(\theta _1 t\right)+2i \xi _1 \tan \left(\theta _1\right) \tan
   \left(\theta _1 t\right)-e^{i \omega _1} \tan ^2\left(\theta _1\right) \sec \left(\theta _1 t\right)}{\xi _1^2-\tan^2\theta_1}\,,\\
\b(t)&=\frac{i \sec \left(\theta _1 t\right) \left(2 \xi _1 \tan \left(\theta _1\right) \sin \left(\omega _1\right)-\xi _1^2 \sin \left(\theta _1
   t\right)-\tan ^2\left(\theta _1\right) \sin \left(\theta _1 t\right)\right)}{\xi _1^2-\tan ^2\left(\theta _1\right)}\\
   \g(t)&=\frac{\xi _2^2 e^{-i \omega _2} \sec \left(\theta _2 t\right)+2i \xi _2 \tan \left(\theta _2\right) \tan
   \left(\theta _2 t\right)-e^{i \omega _2} \tan ^2\left(\theta _2\right) \sec \left(\theta _2 t\right)}{\xi _2^2-\tan^2\theta_2}\,,\\
   \d(t)&=\frac{i \sec \left(\theta _2 t\right) \left(2 \xi _2 \tan \left(\theta _2\right) \sin \left(\omega _2\right)-\xi _2^2 \sin \left(\theta _2
   t\right)-\tan ^2\left(\theta _2\right) \sin \left(\theta _2 t\right)\right)}{\xi _2^2-\tan ^2\left(\theta _2\right)}\,.
\end{split}
\end{align}
The stochastic equation for the wave function is then,
\begin{align}
\begin{split}
\widetilde{\Psi}_{0,\vec{x}}&=\left(e^{i\bar{\w}_1}+\a(t)\right)\widetilde{\Psi}_{0,\vec{x}-\vec{\a}}+\b(t) \widetilde{\Psi}^\star_{0,\vec{x}-\vec{\a}}\,.\\
\widetilde{\Psi}_{1,\vec{x}}&=\left(e^{i\bar{\w}_2}+\g(t)\right)\widetilde{\Psi}_{1,\vec{x}-\vec{\b}}+\d(t) \widetilde{\Psi}^\star_{1,\vec{x}-\vec{\b}}\,.\\
\end{split}
\end{align}
For our specific case, $\widetilde{\Psi}_{\vec{x}}^\star=-\widetilde{\Psi}_{\vec{x}}$. We will use,
\be
\widetilde{\Psi}_{0,\vec{x}}=r_0(\vec{x})\exp[i\chi_0(\vec{x})]\,, \ \widetilde{\Psi}_{1,\vec{x}}=r_1(\vec{x})\exp[i\chi_1(\vec{x})]\,.
\ee
The equations become,
\begin{align}
\begin{split}
\frac{r_0(\vec{x})}{r_0(\vec{x}-\vec{\a})}\exp\left[i(\chi_0(\vec{x})-\chi_0(\vec{x}-\vec{\a}))\right]&=e^{i\bar{\w}_1}+\a(t)-\b(t)\,.\\
\frac{r_1(\vec{x})}{r_1(\vec{x}-\vec{\a})}\exp\left[i(\chi_1(\vec{x})-\chi_1(\vec{x}-\vec{\a}))\right]&=e^{i\bar{\w}_2}+\g(t)-\d(t)
\end{split}
\end{align}
To the leading order in perturbation theory,
\be
\frac{r(\vec{x})}{r(\vec{x}-\vec{\a})}=1+\vec{\a}\cdot\frac{\pd\ln r}{\pd\vec{x}}\,.\ \ \chi(\vec{x})-\chi(\vec{x}-\vec{\a})=1-\vec{\a}\cdot\frac{\pd\chi}{\pd\vec{x}}\,.
\ee
\begin{align}\label{stochasticwave}
\begin{split}
\vec{\a}\cdot\frac{\pd\ln r_0}{\pd\vec{x}}&=\cos\bar{\w}_1-1+\mathfrak{Re}\a(t)\,,\\
\vec{\b}\cdot\frac{\pd\ln r_1}{\pd\vec{x}}&=\cos\bar{\w}_2-1+\mathfrak{Re}\g(t)\,,\\
\end{split}
\end{align}
since $\mathfrak{Re}(\b,\d)=0$. The Lyapunov coefficient is given by,
\be
\frac{1}{N}\sum_{n=1}^N \log |\widetilde{\Psi}_n|=\la\log|\sqrt{r_0^2+r_1^2}|\ra\,.
\ee
For $\mathfrak{Re}\a(t)=\cos\w_1\sec(\theta_1 t)$ and $\mathfrak{Re}\g(t)=\cos\w_2\sec(\theta_2 t)$, the solution to \eqref{stochasticwave} is given by,
\be
r_0^2=\exp\left[2X(\cos(\vec{k}\cdot\vec{\a})-1+\mathfrak{Re}\a(t))\right], r_1^2=\exp\left[2Y(\cos(\vec{k}\cdot\vec{\b})-1+\mathfrak{Re}\g(t))\right]\,.
\ee
Hence,
\begin{align}
\begin{split}
r_0^2+r_1^2=&\exp[2X(\cos(k\cdot\a)-1)]\left(1+\mathfrak{Re}\a(t)+\frac{1}{2}\mathfrak{Re}\a(t)^2+\dots\right)\\
&+\exp[2Y(\cos(k\cdot\b)-1)]\left(1+\mathfrak{Re}\g(t)+\frac{1}{2}\mathfrak{Re}\g(t)^2+\dots\right)\,.
\end{split}
\end{align}
Roughly, by order of estimates, we can assume without generality that,
\be
X(\cos(k\cdot\a)-1)\approx Y(\cos(k\cdot\b)-1)\approx u\,.
\ee
The oscillating terms $\la\mathfrak{Re}\a(t)\ra$, $\la\mathfrak{Re}\g(t)\ra$, average to zero. Hence,
\be
\la\log|\sqrt{r_0^2+r_1^2}|\ra=u/2 \la\left(\mathfrak{Re}\a(t)^2+\mathfrak{Re}\g(t)^2\right)\ra\,.
\ee
which gives an approximate analytical formula for the localization length,
\begin{shaded}
\be
L_{loc}=\frac{2}{\la\left(\mathfrak{Re}\a(t)^2+\mathfrak{Re}\g(t)^2\right)\ra}\,.
\ee
\end{shaded}
upto an overall constant. More precisely, the analytical form of the denominator is given by,
\be
\la\left(\mathfrak{Re}\a(t)^2+\mathfrak{Re}\g(t)^2\right)\ra=\lim_{N\rightarrow\infty}\left(\frac{\cos^2\w_1}{N}\sum_{n=1}^N \sec^2(n \theta_1)+\frac{\cos^2\w_2}{N}\sum_{n=1}^N \sec^2(n \theta_2)\right)\,.
\ee

\section{Conclusions}\label{concl}
In this note, we examine the localization properties of a step dependent coin (SDC) on a random walk. We explore the consequences of SDC walk point wise below:
\begin{itemize}
\item We consider a step dependent coin (SDC) to demonstrate the localization of the wave function in two dimensions. The SDC is based on the same model of coin in \cite{Zahed_2023}. The localization is evident in the probability distribution as well as the boundedness of the number of points with non-zero probability. We also show that by tuning the coin parameter, one can transition through various categories of compact walks ranging  from quantum to classical. As a further check, the return probability to the origin (location of initial wave function) is maximized by certain coins, exhibiting a completely localized walk.

\item In addition to the kinematical evidence, we can also demonstrate the boundedness of the walk from the entropic point of view. The Shannon entropy and entanglement oscillates between zero and an upper bound. The lower bound of zero indicates a separable state which in turn implies a complete localization in the two dimensional grid. The Quantum Relative Entropy ($D$) and Quantum Information Variance ($V$) measures demonstrate the difference between SDC and SIC walks based on the respective density matrices $\r$ and $\s$. For $\theta=\pi/(4j)$, $D(\r||\s)>V(\r||\s)$ and increases monotonically with decreasing $\theta$, while the reverse is true for $\theta=\pi/(4j-1)$. At certain steps, $D(\r||\s)=0$ implying that at those steps, the SDC walk coincides with the SIC walk. While $V(\r||\s)=0$ with $D(\r||\s)\neq0$, implies that the operators for SDC and SIC come arbitrarily close to a maximally entangled state. 
\item In order to characterize the localization phenomenon, we compute the Lyapunov exponent from the transfer matrix method in \cite{Vakulchyk_2017}. The transfer matrix suggests an Anderson Hamiltonian with a non-trivial hopping term. For $\theta=\pi/(4j)$, the localization length is zero which is consistent with a completely localized wave-function. For $\theta=\pi/(4j-1)$, the localization length peaks at $\w\sim\pm \pi/2$. We compare this with an approximate analytic computation of the localization length (where we put in the SDC as a perturbative on a SIC background) and observe the peaks at the same positions. This is similar to the localization dependency on frequency in \cite{Vakulchyk_2017}.
\end{itemize}
However, there are open questions still to be addressed:
\begin{itemize}
\item So far we have restricted ourselves to a step dependent coin in $2d$. A more illustrative generalization would be a site dependent coin as discussed in \cite{Vakulchyk_2017}. A site dependent coin introduces exotic features in the localization features of the Hamiltonian. 
\item A more general computation will be to use a $SU(4)$ coin. However that would introduce 15 parameters which might be difficult to tackle. One way to generalize the computation will be to use a perturbation in place of the zero elements. At this point, the two component wave functions couple with each other perturbatively. 
\item An immediate question would be to address the structure of the Anderson Hamiltonian. The form of the transfer matrix suggests a linear combination of two $1d$ Hamiltonians. However, the forms of the Hamiltonians are different. This Hamiltonian formalism can be used to to analyze phase transitions in various systems including non-Hermitian models of \cite{Luo_2021}. In addition, we would like to compare how and under what circumstances, our step compares with the disordered spin chain systems of \cite{PhysRevA.99.060101,PhysRevE.104.054105} or entangled photons in \cite{Crespi_2013}. 
\item It would be nice to get a quantitative handle on various generalized entanglement measures that characterize the coin operator through entangling power \cite{Zanardi:2000zz, Zahed_2023} or its $n-$qubit generalizations considered in \cite{Badziag_2002, Chen_2005, Lohmayer,Eltschka_2008}. 
\end{itemize}

\section{Acknowledgements}
The author thanks Ahmadullah Zahed, Aranya Bhattacharya and Siddhartha E.M. Guzm\'{a}n for their insightful comments on the work. The author is not aware of any conflict of interest with regards to this work. All the relevant references have been properly marked and due credit given. The data associated with the work has been provided in the figures. Additional data will be provided on request. The author is supported by the S\~{a}o Paulo Funding Agency FAPESP Grants 2021/02304-3 and 2019/24277-8.

\section{Data Availability}
The data relevant to the work is made available in the form of plots in the draft. Additional codes and data will be made available on request.

\bibliography{rw}
\bibliographystyle{utphys}

\end{document}